\documentclass[10pt,aps,prc,amsmath,amssymb,twocolumn,floatfix,nofootinbib,
showpacs,superscriptaddress]{revtex4-1}

\usepackage[utf8]{inputenc}
\usepackage[english]{babel}
\usepackage{amssymb,amsthm,amsmath,amstext,amsbsy,amsopn}
\usepackage{bbm}
\usepackage{url}
\usepackage{nicefrac}
\usepackage{graphicx}
\usepackage{feynmf}
\usepackage{ifpdf}
\usepackage{pstricks,pst-plot}
\usepackage{leftidx}
\usepackage{environ}
\usepackage{suffix}
\usepackage{gensymb}
\usepackage{xspace}

\newcommand{\ie}{\textit{i.e.}}

\newcommand{\cf}{\textit{cf.}\xspace}

\newcommand{\etal}{\textit{et al.}\xspace}

\newcommand{\dd}{\mathrm{d}}

\newcommand{\ii}{\mathrm{i}}

\newcommand{\bra}[1]{\langle #1|}

\newcommand{\ket}[1]{|#1\rangle}

\newcommand{\braket}[2]{\langle #1|#2\rangle}

\newcommand{\mbraket}[3]{\langle #1|#2|#3\rangle}

\NewEnviron{subalign}[1][]{%
\begin{subequations}\begin{align}
  \BODY
\end{align}\label{#1}\end{subequations}
}

\NewEnviron{spliteq}{%
\begin{equation}\begin{split}
  \BODY
\end{split}\end{equation}
}

\newcommand{\MeV}{\ensuremath{\mathrm{MeV}}}
\newcommand{\fm}{\ensuremath{\mathrm{fm}}}

\newcommand{\CG}[6]{\ensuremath{\braket{#1\,#2\,#3\,#4}{#5\,#6}}}
\WithSuffix\newcommand\CG*[6]{\ensuremath{\braket{#1\,#2}{#3\,#4\,#5\,#6}}}

\newcommand{\beq}{\begin{equation}}
\newcommand{\eeq}{\end{equation}}
\newcommand{\beqstar}{\begin{equation*}}
\newcommand{\eeqstar}{\end{equation*}}
\newcommand{\bea}{\begin{eqnarray}}
\newcommand{\eea}{\end{eqnarray}}
\newcommand{\la}{\langle}
\newcommand{\ra}{\rangle}
\newcommand{\bi}{\begin{itemize}}
\newcommand{\ei}{\end{itemize}}

\newcommand{\beal}{\begin{align}}
\newcommand{\eeal}{\end{align}}
\newcommand{\mbf}{\mathbf}
\newcommand{\pp}{p^\prime}
\newcommand{\Ep}{E^\prime}

\newcommand{\GreensFn}{G_0}
\newcommand{\ampT}{\mathcal{T}}
\newcommand{\fL}{f_L}
\newcommand{\thetacm}{\theta^{\prime}}
\newcommand{\phicm}{\varphi^{\prime}}
\newcommand{\thetacprime}{\alpha^\prime}
\newcommand{\thetacdoubleprime}{\alpha^{\prime \prime}}
\newcommand{\msf}{m_{s_f}}
\newcommand{\mJd}{m_{J_d}}
\newcommand{\dcostheta}{\dd \! \cos \theta}
\newcommand{\mbfq}{\mbf{q}}

\begin{document}

\title{Deuteron electrodisintegration with unitarily evolved potentials}

\author{S.~N.~More}
\email{more.13@osu.edu}
\affiliation{Department of Physics, The Ohio State University,
Columbus, Ohio 43210, USA}

\author{S.~König}
\email{koenig.389@osu.edu}
\affiliation{Department of Physics, The Ohio State University,
Columbus, Ohio 43210, USA}

\author{R.~J.~Furnstahl}
\email{furnstahl.1@osu.edu}
\affiliation{Department of Physics, The Ohio State University,
Columbus, Ohio 43210, USA}

\author{K.~Hebeler}
\email{kai.hebeler@physik.tu-darmstadt.de}
\affiliation{Institut für Kernphysik, Technische Universität Darmstadt,
64289 Darmstadt, Germany}
\affiliation{ExtreMe Matter Institute EMMI, GSI Helmholtzzentrum
für Schwerionenforschung GmbH, 64291 Darmstadt, Germany}

\date{\today}

\begin{abstract}

Renormalization group (RG) methods used to soften Hamiltonians for nuclear
many-body calculations change the effective resolution of the interaction.  For
nucleon knock-out processes, these RG transformations leave cross sections
invariant, but initial-state wave functions, interaction currents, and
final-state interactions are individually altered.  This has implications for 
the factorization of nuclear structure and reactions.  We use deuteron
electrodisintegration as a controlled laboratory for studying how structure and 
reaction components are modified under RG evolution, without the complication 
of three-body forces or currents.  The dependence of these changes on 
kinematics is explored.

\end{abstract}


\maketitle

\section{Introduction}
\label{sec:intro}

Softened or ``low momentum'' interactions are widely used in contemporary
nuclear structure calculations because they exhibit faster convergence for
methods using basis expansions (this includes coupled cluster, configuration
interaction, in-medium similarity renormalization group, and self-consistent
Green's function methods)~\cite{Furnstahl:2013oba, Bogner:2009bt}.
Such interactions are derived using unitary transformations starting from
chiral effective field theory or phenomenological interactions that exhibit
significant coupling of high- and low-momentum physics.  When done in small
steps, these unitary transformations are a type of renormalization group (RG)
transformation.  The RG decoupling scale can be associated with the resolution
of the interaction~\cite{Bogner:2009bt}.

But how do we handle observables involving external probes when using
such interactions?  Nuclear structure has conventionally been treated as
largely separate from nuclear reactions.  However, this separation implies a
unique factorization of experimental cross sections into the structure and
reaction parts.  The RG perspective informs us that such a division is itself
inevitably resolution dependent~\cite{Furnstahl:2010wd}.  In some circumstances
the dependence is small and one can define the separation with negligible
ambiguity.  But the significant (and beneficial) changes to wave functions from
evolving to lower resolution with RG methods imply significant changes to this
separation.  That is, what is structure at one resolution becomes part of the
reaction mechanism at another resolution (and vice versa).  This separation is
not only scale dependent, it is \emph{scheme} dependent as well; that is, it
depends on \emph{how} the separation is carried out and on the details of the
original Hamiltonian.

This observation raises questions of consistency and uniqueness in analyzing
and interpreting nuclear experimental data.  For example, it is clear
from previous calculations using the similarity RG (or
SRG)~\cite{Anderson:2010aq,Furnstahl:2013oba,Furnstahl:2013dsa} that the
high-momentum tail of the momentum distribution in a nucleus is dramatically
resolution dependent for the range of decoupling scales used in present-day
nuclear structure calculations.  How then can such a distribution be said to be
extracted from experiment?  Yet it is common in the literature that
high-momentum components are treated as measurable, at least
implicitly~\cite{Frankfurt:2008zv,Arrington:2011xs,Rios:2013zqa,
Boeglin:2015cha}.
In fact, what can be extracted is the momentum distribution at some scale, and
with the specification of a scheme.  This makes momentum distributions model
dependent~\cite{Ford:2014yua, Sammarruca:2015hba}.

This is relevant for recent work to extract a
``nuclear contact''~\cite{Hen:2014lia,Weiss:2014gua,Weiss:2015mba}.
For some physical systems under certain conditions, such as the high-momentum
$1/k^4$ tail in cold atoms at unitarity~\cite{Hoinka:2013fsa}, the scale and
scheme dependence is negligible, so it \emph{can} be determined essentially
uniquely.  But for nuclei this dependence may be substantial and one needs to
carefully define the short-distance content of the nuclear contact.
As illustrated in \cite{Duguet:2014tua}, details of the nuclear shell
structure such as single-particle energies are not measurable.  Their
extraction from experimental data involves fixing a scale and a scheme.

In general, to be consistent between structure and reactions one must calculate
cross sections or decay rates within a single framework.  That is, one must use
the same Hamiltonian and consistent operators throughout the calculation (which
means the same scale and scheme).  Such consistent calculations have existed
for some time for few-body nuclei (e.g.,
see~\cite{Epelbaum:2008ga,Hammer:2012id,Carlson:2014vla,Marcucci:2015rca}) and
are becoming increasingly feasible for heavier nuclei because of advances
in reaction technology, such as using complex basis states to handle continuum
physics.  Recent examples in the literature include
No Core Shell Model Resonating Group Method
(NCSM/RGM)~\cite{Quaglioni:2015via}, coupled cluster~\cite{Bacca:2013dma},
and lattice EFT calculations~\cite{Pine:2013zja}.
But there are many open
questions about constructing consistent currents and how to compare results
from two such calculations.
Some work along this direction was done in \cite{Schuster:2014lga, Neff:2015xda}
where the mean values for eigenstates of renormalized Hamiltonians were
calculated using evolved operators.
We seek to extend this to transition matrix elements and explore the
connection to high-momentum physics in a nucleus in a controlled manner.

In particular, we will take the first steps in exploring the interplay of
structure and reaction as a function of kinematic variables and SRG decoupling
scale $\lambda$ in a controlled calculation of a knock-out process.  There are
various complications for such processes.  With RG evolution, a
Hamiltonian---even with only a two-body potential initially---will develop
many-body components as the decoupling scale decreases.  Similarly, a one-body
current will develop two- and higher-body components.

Our strategy is to avoid dealing with all of these complications simultaneously
by considering the cleanest knock-out process: deuteron electrodisintegration
with only an initial one-body current.  With a two-body system, there are no
three-body forces or three-body currents to contend with.  Yet it still
includes several key ingredients to investigate: i) the wave function will
evolve with changes in resolution; ii) at the same time, the one-body current
develops two-body components, which are simply managed; and iii) there are
final-state interactions (FSI).  It is these ingredients that will mix under
the RG evolution.  We can focus on different effects or isolate parts of the
wave function by choice of kinematics.  For example, we can examine when the
impulse approximation is best and to what extent that is a resolution-dependent
assessment.

We will vary the interaction resolution using SRG transformations, which have
proven to be technically feasible for evolving three-body
forces~\cite{Jurgenson:2009qs,Jurgenson:2010wy,Hebeler:2012pr,Wendt:2013bla}.
The SRG
series of unitary transformations ensures that cross sections are invariant
under changes in resolution.  As the SRG $\lambda$ is varied, the Hamiltonian
$H(\lambda)$ and the nuclear wave function change, the current operator
changes, and the FSI change as well.  The question we address is: How do these
combine to achieve the invariance of the observable cross section?

The electron scattering knock-out process is particularly interesting because
of the connection to past, present, and planned
experiments~\cite{Boffi:1996, LENP_white_paper2015}.
The conditions for clean factorization
of structure and reactions in this context is closely related to the impact of
3N forces,
two-body currents, and final-state interactions, which have not been cleanly
understood as yet~\cite{Furnstahl:2010wd}.
All of this becomes particularly relevant for high-momentum-transfer electron
scattering.%
\footnote{Note that high-momentum transfers imply high-resolution
\emph{probes}, which is different from the resolution induced by the RG
transformation decoupling scale.  How the latter should be chosen to best
accommodate the former is a key unanswered question.}
This physics is conventionally explained in terms of short-range correlation
(SRC) phenomenology~\cite{Frankfurt:2008zv,Atti:2015eda}.  SRCs are two- or
higher-body components of the nuclear wave function with high relative momentum
and low center-of-mass momentum.  These explanations would seem to present a
puzzle for descriptions of nuclei with low-momentum Hamiltonians, for which
SRCs are essentially absent from the wave functions.

This puzzle is resolved by the unitary transformations that mandate the
invariance of cross section.  The physics that was described by SRCs in the
wave functions must shift to a different component, such as a two-body
contribution from the current.  This may appear to complicate the reaction
problem just as we have simplified the structure part, but past work and
analogies to other processes suggests that factorization may in fact
become cleaner~\cite{Anderson:2010aq,Bogner:2012zm}.  One of our goals is to
elucidate this issue, although we will only start to do so in the present work.

The underlying picture is analogous to that used in deep inelastic scattering
(DIS)~\cite{Furnstahl:2010wd,Furnstahl:2013dsa}.  In the analysis of DIS, one
introduces both a renormalization and a factorization scale.  In the usual
dimensional regularization minimal subtraction scheme $\overline{\rm MS}$, the
renormalization scale $\mu_R$ sets the division between long- and
short-distance physics in the Hamiltonian.  This is manifested in the running
coupling $\alpha_s(\mu)$ where the choice of $\mu = \mu_R$ is to optimize the
efficacy of the QCD perturbation expansion.  The factorization scale $\mu_F$
dictates the division between what goes into the reaction part, namely the
purely hard process described in pQCD, and the structure part, which is
subsumed into the soft parton distribution functions.  Changing $\mu_F$ changes
the balance.  In many cases these two scales are chosen to be the same and
equal to the magnitude of the four-momentum transfer $Q$, in order to minimize
the contribution of logarithms that can disturb the perturbative expansion.

The nuclear analogs to these two scales are tied together in the SRG evolution.
In particular, the decoupling scale, roughly given by the value of the SRG flow
parameter $\lambda$, sets both these scales.  For the Hamiltonian, this scale
clearly sets the division between low and high momentum.  For operators acting
on wave functions, the decoupling dictates the division between the two; e.g.,
the scale at which a one-body current is largely replaced by a two-body
current.  In the present paper, we will illustrate the combined interplay with
quantitative calculations.  We build upon previous work by Anderson~\etal on
SRG operator evolution for the deuteron~\cite{Anderson:2010aq}, and the work by
Yang and Phillips~\cite{Yang:2013rza} in applying chiral EFT to deuteron
electrodisintegration.

This paper is organized as follows.  In Sec.~\ref{sec:formalism}, we briefly
review the electrodisintegration formalism and develop the machinery we need for
the SRG evolution, which we accomplish in practice by appropriate insertions of
unitary transformation matrices.  In Sec.~\ref{sec:results}, we present
proof-of-principle tests and illustrate the interplay
of the different components that enter in the calculation of the disintegration
process.  We give representative results for selected kinematics.  In
Sec.~\ref{sec:conclusion}, we summarize our observations and plans to extend the
calculations to other kinematics, as well as beyond the deuteron system and
one-body initial currents.

\section{Formalism}
\label{sec:formalism}

\subsection{Deuteron electrodisintegration: a primer}
\label{subsec:disintegration_primer}

Deuteron electrodisintegration is the simplest nucleon-knockout process
and has been considered as a test ground for various $NN$ models for a
long time (see, for example, Refs.~\cite{Arenhovel:2004bc,Boeglin:2015cha}).
It has also been well studied experimentally~\cite{Gilad:1998wia,Egiyan:2007qj}.
As outlined in the introduction, the absence of three-body currents and forces
makes it an ideal starting point for studying the interplay with SRG evolution
of the deuteron wave function, current, and final-state interactions.

\begin{figure}[htbp]
 \centering
 \includegraphics[width=0.975\columnwidth]{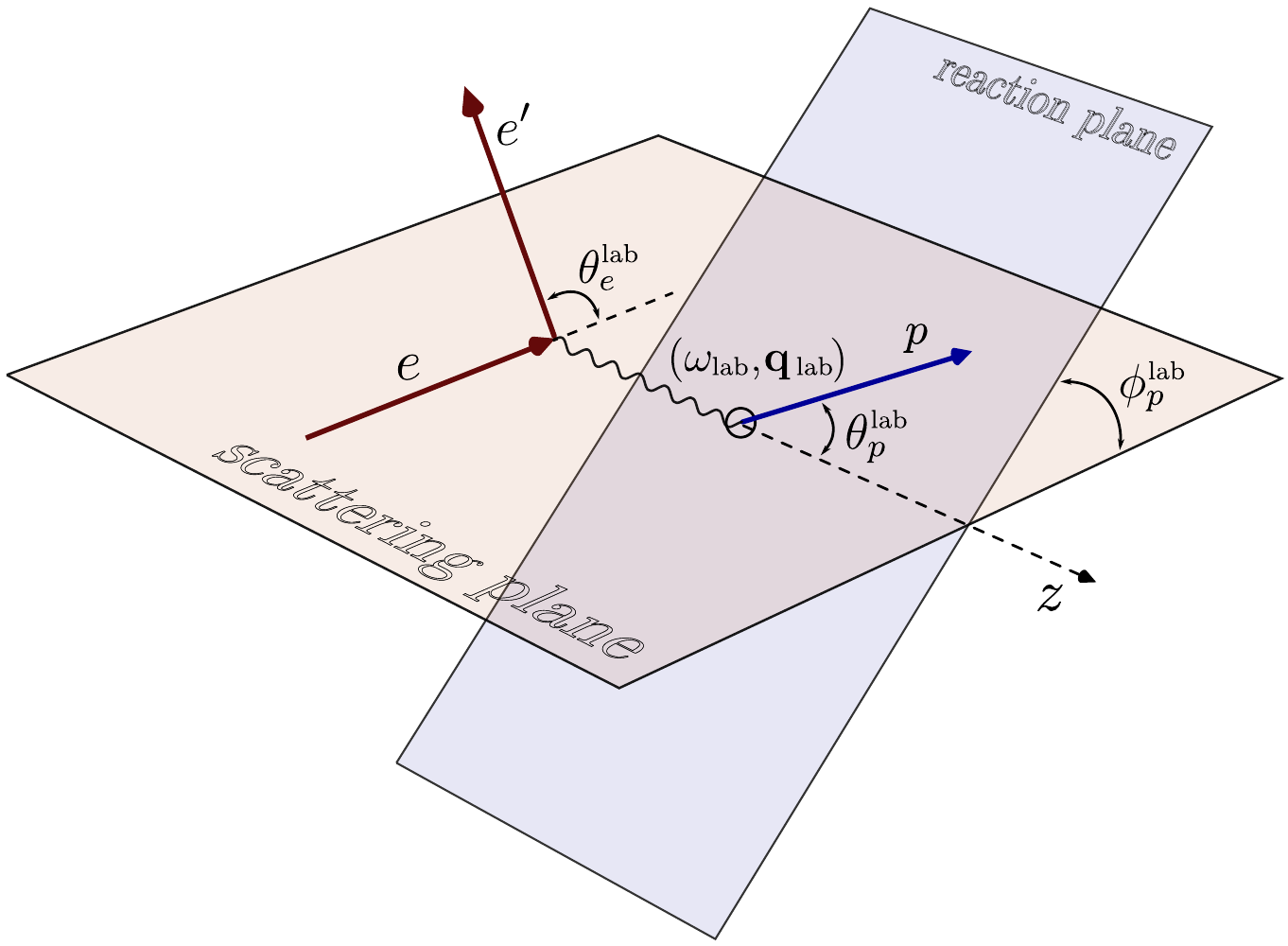}
 \caption{(color online) The geometry of the electro-disintegration process in
 the lab frame.  The virtual photon disassociates the deuteron into the proton
 and the neutron (not shown in this figure).}
 \label{fig:deut_dis_kinematics}
\end{figure}

We follow the approach of Ref.~\cite{Yang:2013rza}, which we briefly review.
The kinematics for the process in the laboratory frame is shown in
Fig.~\ref{fig:deut_dis_kinematics}.
The virtual photon from electron scattering transfers enough energy and momentum
to break up the deuteron into a proton and neutron.  The differential cross
section for deuteron electrodisintegration for unpolarized scattering in the lab
frame is given by~\cite{Arenhovel:1988qh}
\begin{multline}
 \frac{\dd^3 \sigma}{\dd {k^\prime}^{\rm lab} \dd\Omega_e^{\rm lab}
 \dd \Omega_p^{\rm lab}}
 = \frac{\alpha}{6 \, \pi^2} \frac{{k^\prime}^{\rm lab}}
 {k^{\rm lab} (Q^2)^2}
 \Big[
  v_L \fL + v_T f_T \\
  \null + v_{TT} f_{TT} \cos 2 \phi_p^{\rm lab}
  + v_{LT} f_{LT} \cos \phi_p^{\rm lab}
 \Big] \,.
\end{multline}
Here $\Omega_e^{\rm lab}$ and $\Omega_p^{\rm lab}$ are the solid angles of the
electron and the proton, $k^{\rm lab}$ and ${k^\prime}^{\rm lab}$ are the
magnitude of incoming and outgoing electron 3-momenta,  $Q^2$ is the
4-momentum-squared of the virtual photon, and $\alpha$ is the fine structure
constant.  $\phi_p^{\rm lab}$ is the angle between the scattering plane
containing the electrons and the plane spanned by outgoing nucleons.
$v_L\,, v_T\,, \ldots$ are electron kinematic factors, and $\fL, f_T, \ldots$
are the deuteron structure functions.  These structure functions contain all
the dynamic information about the process.
The four structure functions are independent and can be separated by combining
cross-section measurements carried out with appropriate kinematic
settings~\cite{Kasdorp:1997ba}.  Structure functions are thus cross sections up
to kinematic factors and are independent of the SRG scale $\lambda$.  In this
work we focus on the longitudinal structure function $\fL$, following the
approach of Ref.~\cite{Yang:2013rza}.

\subsection{Calculating $\fL$}
\label{sec:calculating_fl}

As in Ref.~\cite{Yang:2013rza}, we carry out the calculations in
the center-of-mass frame of the outgoing proton-neutron pair.
In this frame the photon
four-momentum is $(\omega,\mbf{q})$, which can be obtained from the
initial electron energy and $\theta_e$, the electron scattering angle.
We denote the momentum of the outgoing proton by $\mathbf{\pp}$ and
take $\mbf{q}$ to be along $z$-axis.  The angles of $\mbf{\pp}$ are
denoted by $\Omega_{\mbf{\pp}} = (\thetacm, \phicm)$.

The longitudinal structure function can be written as
\begin{equation}
 \fL = \sum_{\substack{S_f, m_{s_f}\\ \mJd}}
 \ampT_{S_f,m_{s_f},\mu = 0,\mJd}\!(\thetacm, \phicm) \,
 \ampT_{S_f, m_{s_f},\mu = 0,\mJd}^\ast\!(\thetacm, \phicm) \,,
\label{eq:f_L_from_T}
\end{equation}
where $S_f$ and $m_{s_f}$ are the spin quantum numbers of the final
neutron-proton state, $\mu$ is the polarization index of the virtual photon,
and $\mJd$ is the angular momentum of the initial deuteron
state. The amplitude $\ampT$ is given by~\cite{Arenhoevel:1992xu}
\begin{equation}
 \ampT_{S,\msf,\mu,\mJd}
 = -\pi \sqrt{2\alpha|\mathbf{\pp}|E_p E_d / M_d}
 \,\la \psi_{f} \,|\, J_{\mu}(\mathbf{q}) \,| \psi_i \ra \,,
\label{eq:T_definiton}
\end{equation}
where $\bra{\psi_f}$ is the final-state wavefunction of the outgoing
neutron-proton pair, $\ket{\psi_i}$ is the initial deuteron state, and
$J_{\mu} (\mbf{q})$ is the current operator that describes the momentum
transferred by the photon.
The variables in Eq.~\eqref{eq:T_definiton} are:
\begin{itemize}
\item fine-structure constant $\alpha$;
\item outgoing proton (neutron) 3-momentum $\mbf{\pp} \, (-\mbf{\pp})$;
\item proton energy $\displaystyle E_p = \sqrt{M^2 + \mbf{\pp}^2}$,
where $M$ is the average of proton and neutron mass
\item deuteron energy $\displaystyle E_d = \sqrt{M_d^2 + \mbf{q}^2} $,
where $M_d$ is the mass of the deuteron.
\end{itemize}
As mentioned before, all of these quantities are in the center-of-mass
frame of the outgoing nucleons.

For $f_L$, $\mu = 0$ and therefore only $J_0$ contributes.
The one-body current matrix element is given by
\begin{multline}
 \la \mbf{k}_1 \, T_1| \, J_0(\mbf{q}) \, | \,\mbf{k}_2 \, T=0 \ra \\
 = \frac{1}{2} \big(G_E^p + (-1)^{T_1} G_E^n\big) \,
 \delta(\mbf{k}_1 - \mbf{k}_2 - \mbf{q}/2) \\
 \null + \frac{1}{2} \big((-1)^{T_1} G_E^p +  G_E^n\big) \, \delta(\mbf{k}_1
 - \mbf{k}_2 + \mbf{q}/2) \,,
\label{eq:J0_def}
\end{multline}
where $G_E^p$ and $G_E^n$ are the electric form factors of the proton and the
neutron, and the deuteron state has isospin $T=0$.

The final-state wave function of the outgoing proton-neutron pair can be written
as
\begin{equation}
 |\psi_f\ra = | \phi \ra + G_0 (E^\prime) \, t(E^\prime) \,| \phi \ra \,,
\label{eq:psi_f_def}
\end{equation}
where $\ket{\phi}$ denotes a relative plane wave, $\GreensFn$ and $t$
are the Green's function and the $t$-matrix respectively, and  $E^\prime =
\mbf{\pp}^2/M$ is the energy of the outgoing nucleons.  The second term in
Eq.~\eqref{eq:psi_f_def} describes the interaction between the outgoing
nucleons.

In the impulse approximation (IA) as defined here, the interaction between the
outgoing nucleons is ignored and $|\psi_f \ra_{\rm IA} \equiv | \phi \ra$.
The plane wave $|\phi\ra$ will have both isospin $0$ and $1$ components.
The current $J_0$, $\GreensFn$, and the $t$-matrix are diagonal in spin space.
The deuteron has spin $S=1$ and therefore the final state will also have $S=1$.
Hence, we have
\begin{spliteq}
 \ket{\phi} &\equiv \ket{\mbf{\pp}\, S=1 \, \msf \psi_T} \\
 &= \frac{1}{2} \sum_{T = 0,1}
 \Big(\ket{\mbf{\pp} \, S=1 \, \msf} \\[-0.5em]
 &\hspace{6em}\null + (-1)^T \ket{{-}\mbf{\pp} \, S=1 \, \msf} \Big) \,\ket{T}
 \,.
\label{eq:phi_def}
\end{spliteq}
Using Eqs.~\eqref{eq:J0_def} and~\eqref{eq:phi_def}, the overlap matrix element
in IA becomes
\begin{widetext}
\begin{multline}
 \la \psi_f | \, J_0 \, |\psi_i \ra_{\rm IA}
 = \sqrt{\frac{2}{\pi}} \sum_{L_{d} = 0, 2}
 \CG{L_d}{\mJd - \msf}{S=1}{\msf}{J=1}{\mJd} \\
 \null \times \Big[
  G_E^p \, \psi_{L_d}(|\mbf{\pp} - \mbf{q}/2|)
  \,Y_{L_d,\mJd-m_{s_f}}\!(\Omega_{\mbf{\pp} - \mbf{q}/2})
  + G_E^n \, \psi_{L_d}(|\mbf{\pp} + \mbf{q}/2|)
  \,Y_{L_d,\mJd-m_{s_f}}\!(\Omega_{\mbf{\pp} + \mbf{q}/2})
 \Big] \,,
\label{eq:overlap_IA}
\end{multline}
\end{widetext}
where $\Omega_{\mbf{\pp} \pm \mbf{q}/2}$ is the solid angle between the
unit vector $\hat{z}$ and $\mbf{\pp}\pm\mbf{q}/2$.  $\psi_{L_d}$ is the deuteron
wave function in momentum space defined as
\begin{multline}
 \braket{k_1 \, J_1 \, m_{J_1} \, L_1 \, S_1 \, T_1}{\psi_i}
 = \psi_{L_1}(k_1) \\
 \null \times \delta_{J_1,1}\delta_{m_{J_1},\mJd}
 \delta_{L_1,L_d}\delta_{S_1,1}\delta_{T_1,0} \,.
\end{multline}
The $S$-wave $(L=0)$ and $D$-wave $(L=2)$ components of the deuteron wave
function satisfy the normalization condition
\begin{equation}
 \frac{2}{\pi} \int dp \, p^2 \, \big(\psi_0^2(p) + \psi_2^2(p)\big) = 1 \,.
\end{equation}
In deriving Eq.~\eqref{eq:overlap_IA} we have used the property of the
spherical harmonics that
\begin{equation}
 Y_{lm}(\pi-\theta,\phi+\pi) = (-1)^l \, Y_{lm}(\theta,\phi) \,.
\end{equation}
In our work we follow the conventions of Ref.~\cite{Landau:1989}.  Since
$\thetacm$ and $\phicm$ are the angles of $\mbf{\pp}$, $\Omega_{\mbf{\pp} -
\mbf{q}/2} \equiv \big(\thetacprime(\pp, \thetacm ), \phicm \big)$
and $\Omega_{\mbf{\pp} + \mbf{q}/2} \equiv
\big(\thetacdoubleprime(\pp, \thetacm),
\phicm \big)$, where
\begin{equation}
 \thetacprime (\pp, \thetacm) = \cos^{-1}
 \!\left(
  \frac{\pp \cos \thetacm - q/2}{\sqrt{{\pp}^2 - \pp q \cos\thetacm + q^2/4}}
 \right)
\label{eq:theta_c_prime_def}
\end{equation}
and
\begin{equation}
 \thetacdoubleprime (\pp, \thetacm) = \cos^{-1}
 \!\left(
  \frac{\pp \cos \thetacm + q/2}{\sqrt{{\pp}^2 + \pp q \cos\thetacm + q^2/4}}
 \right) \,.
\label{eq:theta_c_double_prime_def}
\end{equation}

The overlap matrix element including the final-state interactions (FSI) is
given by
\begin{equation}
 \la \psi_f| \, J_0 \,|\psi_i \ra
 = \underbrace{\la \phi| \, J_0 \,|\psi_i \ra}_{\rm IA}
 + \underbrace{\la \phi| t^\dag \, \GreensFn^\dag\, J_0 \,|\psi_i \ra}_{\rm FSI}
 \,.
\label{eq:overlap_IA_plus_FSI}
\end{equation}
The first term on the right side of Eq.~\eqref{eq:overlap_IA_plus_FSI}
has already been evaluated in Eq.~\eqref{eq:overlap_IA}.  Therefore, the term
we still need to evaluate is $\la \phi| t^\dag \, \GreensFn^\dag\, J_0
\,|\psi_i \ra$.  The $t$-matrix is most conveniently calculated in a
partial-wave basis.  Hence, the FSI term is evaluated by inserting complete
sets of states in the form
\begin{equation}
 1 = \frac{2}{\pi} \sum_{\substack{L, S \\ J,  m_J }}
 \sum_{T = 0, 1} \int dp \,p^2 \, |p \, J \, m_J \, L \, S \, T \ra
 \, \la p \, J\, m_J \, L \, S \, T | \,.
\label{eq:completeness_partial_wave}
\end{equation}
The outgoing plane-wave state in the partial-wave basis is given by
\begin{multline}
 \la \phi | \, k_1 \, J_1 \, m_{J_1} \, L_1 \, S = 1 \, T_1 \ra
  =  \frac{1}{2} \, {\sqrt\frac{2}{\pi}} \, \frac{\pi}{2} \,
 \frac{\delta(\pp - k_1)}{k_1 ^2}
  \\
  \null\times
 \CG{L_1}{m_{J_1} - \msf}{S=1}{\msf}{J_1}{m_{J_1}}
 \, \\
 \null \times \big(1 + (-1)^{T_1} (-1)^{L_1}\big)
 \,Y_{L_1,m_{J_1} - \msf}\!(\thetacm, \phicm) \,.
\label{eq:phi_def_pw}
\end{multline}
The Green's
function is diagonal in $J$, $m_J$, $L$, $S$, and $T$, so we have
\begin{equation}
 \la k_1 | \, \GreensFn^\dag \,| k_2 \ra
 = \frac{\pi}{2} \, \frac{\delta(k_1 - k_2)}{k_1^2}
 \, \frac{M}{{\pp}^2 - k_1 ^2 - i\epsilon} \,.
\label{eq:G0_def_pw}
\end{equation}

We also need to express the current in Eq.~\eqref{eq:J0_def} in the partial-wave
basis.  To begin with, let us just work with first term in
Eq.~\eqref{eq:J0_def}, which we denote by $J_0 ^-$.  In the partial-wave basis,
it is written as
\begin{widetext}
\begin{multline}
 \mbraket{k_1 \, J_1 \, \mJd \, L_1 \, S = 1 \, T_1}{J_0^{-}}
 {\,k_2 \, J=1 \, \mJd \, L_2 \, S = 1 \, T = 0}
 = \frac{\pi^2}{2} \, \big(G_E ^p + (-1)^{T_1} \, G_E^n\big) \\
 \times \sum_{\widetilde{m}_s = -1}^1 \int\!\dcostheta
 \,\CG*{J_1}{\mJd}{L_1}{\mJd - \widetilde{m}_s}{S = 1}{\widetilde{m}_s}
 \,P_{L_1}^{\mJd - \widetilde{m}_s}\!(\cos\theta)
 \,P_{L_2}^{\mJd - \widetilde{m}_s}\!\big(\cos\thetacprime(k_1, \theta)\big) \\
 \null\times \frac{\delta\big(k_2-\sqrt{k_1^2-k_1 q \cos\theta + q^2/4}\big)}
 {k_2^2}
 \,\CG{L_2}{\mJd - \widetilde{m}_s}{S = 1}{\widetilde{m}_s}{J = 1}{\mJd} \,.
\label{eq:J_0_minus_def_pw}
\end{multline}
Here $\mJd$ is the deuteron quantum number, which is preserved throughout.
We have used the deuteron quantum numbers in the ket in anticipation
that we will always evaluate the matrix element of $J_0$ with the deuteron
wave function on the right.   $\thetacprime$ is as defined in
Eq.~\eqref{eq:theta_c_prime_def}.  In deriving Eq.~\eqref{eq:J_0_minus_def_pw}
we have also made use of the relation \cite{Jerry_thesis}
\begin{equation}
 \int Y_{lm} ^\ast(\theta, \varphi) \,
 Y_{l^\prime m^\prime}(\alpha^\prime,\varphi)
 \,\dd\!\cos \theta \, \dd \varphi
 = 2\pi \delta_{m m^\prime}
 \int\!\dcostheta
 P_l^m(\cos\theta) \, P_{l^\prime}^m(\cos \alpha^\prime) \,.
\end{equation}

Equations~\eqref{eq:phi_def_pw}, \eqref{eq:G0_def_pw},
and~\eqref{eq:J_0_minus_def_pw} can be combined to obtain
\begin{multline}
 \mbraket{\phi}{t^\dag \, \GreensFn^\dag \, J_0^-}{\psi_i}
 = \sqrt{\frac{2}{\pi}} \,\frac{M}{\hbar c}
 \sum_{T_1 = 0,1} \big(G_E ^p + (-1)^{T_1} \, G_E^n\big)
 \sum_{L_1 = 0}^{L_{\rm max}} \big(1 + (-1)^{T_1} (-1)^{L_1}\big)
 \, Y_{L_1 , \mJd - \msf}\!(\thetacm, \phicm) \\
 \null\times \sum_{J_1 = |L_1 - 1|}^{L+1}
 \CG{L_1}{\mJd - \msf}{S=1}{\msf}{J_1}{\mJd}
 \sum_{L_2 = 0}^{L_{\rm max}} \int \dd k_2 \, k_2^2 \,
  t^\ast(k_2, \pp, L_2, L_1, J_1, S = 1, T_1) \\
 \null\times \sum_{\widetilde{m}_s = -1}^1
 \CG*{J_1}{\mJd}{L_2}{\mJd - \widetilde{m}_s}{S = 1}{\widetilde{m}_s}
 \sum_{L_d = 0,2}
 \CG{L_d}{\mJd - \widetilde{m}_s}{S = 1}{\widetilde{m}_s}{J = 1}{\mJd} \\
 \null\times \int \dcostheta
 \, \frac{1}{{\pp}^2 - k_2^2 - i \epsilon}
 P_{L_2}^{\mJd - \widetilde{m}_s}(\cos\theta) \,
 P_{L_d}^{\mJd - \widetilde{m}_s}\!\big(\cos\thetacprime(k_2,\theta)\big) \,
 \psi_{L_d}\Big(\!\sqrt{{k_2}^2 - k_2 \, q \, \cos\theta + q^2/4}\Big) \,.
\label{eq:phi_t_g0_J0_minus}
\end{multline}
\end{widetext}

We denote the second term in the one-body current Eq.~\eqref{eq:J0_def} by
$J_0^+$.  The expression for $\mbraket{\phi}{t^\dag \, \GreensFn^\dag \,
J_0^+}{\psi_i}$ is analogous to Eq.~\eqref{eq:phi_t_g0_J0_minus}, the only
differences being that the form-factor coefficient is $(-1)^{T_1} G_E^p +
G_E^n$ and the input arguments for the second associated Legendre polynomial
and the deuteron wave function are different.  The two factors respectively
become $\displaystyle P_{L_d}^{\mJd - \widetilde{m}_s}\!
\big(\cos\thetacdoubleprime(k_2, \theta)\big)$
and $\psi_{L_d}\big(\sqrt{{k_2}^2 + k_2 \,q \, \cos\theta + q^2/4}\big)$, where
$\thetacdoubleprime$ is defined in Eq.~\eqref{eq:theta_c_double_prime_def}.
It can be shown that $\mbraket{\phi}{t^\dag \, \GreensFn^\dag \, J_0^+}{\psi_i}
= \mbraket{\phi}{t^\dag \, \GreensFn^\dag \, J_0^-}{\psi_i}$.  Thus,
\begin{equation}
 \mbraket{\phi}{t^\dag \, \GreensFn^\dag \, J_0}{\psi_i}
 = 2 \, \mbraket{\phi}{t^\dag \, \GreensFn^\dag \, J_0^-}{\psi_i} \,.
\label{eq:J0_minus_twice_relation}
\end{equation}
Using this we can evaluate the overlap matrix element in
Eq.~\eqref{eq:overlap_IA_plus_FSI}.  As outlined in Eqs.~\eqref{eq:f_L_from_T}
and \eqref{eq:T_definiton}, this matrix element is related to the
longitudinal structure function $\fL$.  Recall that the deuteron spin is
conserved throughout and therefore $S_f = 1$ in Eq.~\eqref{eq:f_L_from_T}.

In Sec.~\ref{subsec:illustrative_examples} we present results for $\fL$ both in
the IA and including the FSI.  These results match those of
Ref.~\cite{Yang:2013rza,Arenhoevel:1992xu}, verifying the accuracy of the
calculations presented above.

\subsection{Evolution setup}
\label{sec:evolution_set_up}

As outlined in the introduction, we want to investigate the effect of
unitary transformations on calculations of $\fL$.
Let us start by looking at the IA matrix element:
\begin{spliteq}
 \la \phi | J_0 | \psi_i \ra
 &= \la \phi | U^\dag \, U \, J_0 \, U^\dag \, U \, | \psi_i \ra \\
 &= \underbrace{\la\phi|\widetilde{U}^\dag J_0^\lambda|\psi_i^\lambda\ra}_{A}
 + \underbrace{\la \phi | \, J_0^\lambda | \psi_i^\lambda \ra}_{B} \,,
\label{eq:A_B_split_up}
\end{spliteq}
where we decompose the unitary matrix $U$ into the identity and a residual
$\widetilde{U}$,
\begin{equation}
 U = I + \widetilde{U} \,.
\label{eq:U_decomposition}
\end{equation}
The matrix $\widetilde{U}$ is smooth and therefore amenable to interpolation.
The $U$ matrix is calculated following the approach in \cite{Anderson:2010aq}.
The terms in Eq.~\eqref{eq:A_B_split_up} can be further split into

\begin{multline}
 \la \phi | \, J_0^\lambda | \psi_i^\lambda \ra
 = \underbrace{\la \phi | \widetilde{U}\, J_0 \, \widetilde{U}^\dag |
  \psi_i^\lambda \ra}_{B_1}
 + \underbrace{\la \phi | \widetilde{U}\, J_0 \, | \psi_i^\lambda \ra}_{B_2} \\
 \null + \underbrace{\la \phi | \, J_0 \, \widetilde{U}^\dag | \psi_i^\lambda
  \ra}_{B_3}
 + \underbrace{\la \phi |\, J_0 \, | \psi_i^\lambda \ra}_{B_4}
\label{eq:B_split_up}
\end{multline}
and
\begin{multline}
 \la \phi | \widetilde{U}^\dag \, J_0^\lambda | \psi_i^\lambda \ra \\[0.25em]
 = \underbrace{\la \phi | \widetilde{U}^\dag \, \widetilde{U}\, J_0 \,
  \widetilde{U}^\dag | \psi_i^\lambda \ra}_{A_1}
 + \underbrace{\la \phi | \widetilde{U}^\dag \, \widetilde{U}\, J_0 |
  \psi_i^\lambda \ra}_{A_2} \\
 \null + \underbrace{\la \phi | \widetilde{U}^\dag J_0 \,
  \widetilde{U}^\dag | \psi_i^\lambda \ra}_{A_3}
 + \underbrace{\la\phi | \widetilde{U}^\dag J_0 | \psi_i^\lambda\ra}_{A_4} \,.
\label{eq:A_split_up}
\end{multline}
The $B_4$ term is the same as in Eq.~\eqref{eq:overlap_IA}, but with the
deuteron wave function replaced by the evolved version $\psi_{L_d}^\lambda$.
Inserting complete sets of partial-wave basis states as in
Eq.~\eqref{eq:completeness_partial_wave} and using Eqs.~\eqref{eq:phi_def_pw}
and \eqref{eq:J_0_minus_def_pw}, we can obtain the expressions for $B_1$, $B_2$,
$B_3$ and $A_1,\ldots,A_4$.  These expressions are given in
Appendix~\ref{appendix:evol_eqns}.

Using the expressions for $A_1,\ldots, A_4$ and $B_1,\ldots, B_4$, we can obtain
results for $\fL$ in the IA with one or more components of the overlap matrix
element $\mbraket{\phi}{J_0}{\psi}$ evolved.  When calculated in IA, $\fL$ with
all components evolved matches its unevolved counterpart, as shown later
in Sec.~\ref{subsec:illustrative_examples}.
The robust agreement between the evolved and unevolved answers indicates that
the expressions derived for $A_1, \ldots, B_4$ are correct and that there is
no error in generating the $U$-matrices.  In
Sec.~\ref{subsec:numerical_implementation} we provide some details about the
numerical implementation of the equations presented here.

Let us now take into account the FSI and study the effects
of evolution.  The overlap matrix element should again be unchanged under
evolution,
\begin{equation}
 \mbraket{\psi_f}{J_0}{\psi_i}
 = \mbraket{\psi_f^\lambda}{J_0^\lambda}{\psi_i^\lambda} \,,
\end{equation}
where $\psi_f$ is given by Eq.~\eqref{eq:psi_f_def}.  Furthermore,
\begin{equation}
 |\psi_f ^\lambda \ra = | \phi \ra + G_0 \, t_\lambda |\phi \ra \,,
\label{eq:psi_f_lam_def}
\end{equation}
where $t_\lambda$ is the evolved $t$-matrix, \ie, the $t$-matrix obtained
by solving the Lippmann--Schwinger equation using the evolved potential, as
discussed in Appendix~\ref{sec:finalstate}.  Thus
\begin{equation}
 \mbraket{\psi_f^\lambda}{J_0^\lambda}{\psi_i^\lambda}
 = \underbrace{\mbraket{\phi}{J_0^\lambda}{\psi_i^\lambda}}_{B}
 + \underbrace{\mbraket{\phi}{t_\lambda ^\dag \, G_0^\dag
  \, J_0^\lambda}{\psi_i^\lambda}}_{F} \,.
\end{equation}
The term $B$ is the same that we already encountered in
Eq.~\eqref{eq:A_B_split_up}.  The term $F$ can also be split up into four terms:
\begin{multline}
 \mbraket{\phi}{t_\lambda^\dag\,G_0^\dag\,J_0^\lambda}{\psi_i^\lambda}\\[0.25em]
 = \underbrace{\mbraket{\phi}{t_\lambda ^\dag \, G_0^\dag \, \widetilde{U}
  \, J_0 \, \widetilde{U}^\dag}{\psi_i^\lambda}}_{F_1}
 + \underbrace{\mbraket{\phi}{t_\lambda ^\dag \, G_0^\dag \, \widetilde{U}
  \, J_0 }{\psi_i^\lambda}}_{F_2} \\
 \null + \underbrace{\mbraket{\phi}{t_\lambda ^\dag \, G_0^\dag
  \, J_0 \, \widetilde{U}^\dag}{\psi_i^\lambda}}_{F_3}
 + \underbrace{\mbraket{\phi}{t_\lambda ^\dag \, G_0^\dag
  \, J_0}{\psi_i^\lambda}}_{F_4} \,.
\label{eq:F_split_up}
\end{multline}
The expression for $F_4$ can easily be obtained from
Eqs.~\eqref{eq:phi_t_g0_J0_minus} and~\eqref{eq:J0_minus_twice_relation} by
replacing the deuteron wave function and the $t$-matrix by their evolved
counterparts.  As before, we insert complete sets of partial-wave basis states
 using Eq.~\eqref{eq:completeness_partial_wave} and evaluate $F_3$, $F_2$,
and $F_1$; see Eqs.~\eqref{eq:F3}, \eqref{eq:F2}, and \eqref{eq:F1}.
Figures in
Sec.~\ref{subsec:illustrative_examples} compare $\fL$ calculated
from the matrix element with all components evolved to the unevolved $\fL$.
We find an excellent
agreement, validating the expressions for $F_1, \ldots, F_4$.

\subsection{Numerical implementation}
\label{subsec:numerical_implementation}

There are various practical issues in the calculation of evolved matrix elements
that are worth detailing.  We use C++11 for our numerical implementation of
the expressions discussed in the previous section.  Matrix elements with a
significant
number of components evolved are computationally quite expensive due to a large
number of nested sums and integrals (see in particular
Appendix~\ref{appendix:evol_eqns}).

The deuteron wave function and $NN$ $t$-matrix are obtained by discretizing the
Schrödinger and Lippmann--Schwinger equations, respectively; these equations
are also used to interpolate the $t$-matrix and wave function to points not on
the discretized mesh.  For example, if we write the momentum-space Schrödinger
equation---neglecting channel coupling here for simplicity---as
\begin{multline}
 \psi(p) = \int \dd q\,q^2\,G_0(-E_B,q) V(p,q)\,\psi(q) \\
 \rightarrow \sum_{i} w_i\,q_i^2\,G_0(-E_B,q_i) V(p,q_i)\,\psi(q_i) \,,
\label{eq:SG-simple}
\end{multline}
it can be solved numerically as a simple matrix equation by setting
$p\in\{q_i\}$.  For any $p=p_0$ not on this mesh, the sum in
Eq.~\eqref{eq:SG-simple} can then be evaluated to get $\psi(p_0)$.  This
technique is based on what has been introduced in connection with
contour-deformation methods in break-up scattering
calculations~\cite{Hetherington:1965zza,Schmid:1974}.

To interpolate the potential, which is stored on a momentum-space grid, we use
the  two-dimensional cubic spline algorithm from ALGLIB~\cite{ALGLIB:0915}.  In
order to avoid unnecessary recalculation of expensive quantities---in particular
of the off-shell $t$-matrix---while still maintaining an implementation very
close to the expressions given in this paper, we make use of transparent caching
techniques.\footnote{This means that the expensive calculation is only carried
out once, the first time the corresponding function is called for a given set of
arguments, while subsequent calls with the same arguments return the result
directly, using a fast lookup.  All this is done without the \emph{calling} code
being aware of the caching details.}  For most integrations, in particular those
involving a  principal value, we use straightforward nested Gaussian quadrature
rules; only in a few cases did we find it more efficient to use adaptive
routines for multi-dimensional integrals.

With these optimizations, the calculations can in principle still be run on a
typical laptop computer.  In practice, we find it more convenient to use a
small cluster, with parallelization implemented using the TBB
library~\cite{TBB:0915}.  On a node with 48 cores, generating data for a
meaningful plot (like those shown in Sec.~\ref{sec:results}) can then be done
in less than an hour.  For higher resolution and accuracy, we used longer runs
with a larger number of data and integration mesh points.

\section{Effects of unitary evolution}
\label{sec:results}

\subsection{First order analytical calculation}

Recall that from Eqs.~\eqref{eq:f_L_from_T} and~\eqref{eq:T_definiton} we have
\begin{equation}
 \fL \propto \sum_{\msf, \mJd}\left|\mbraket{\psi_f}{J_0}{\psi_i}\right|^2 \,.
\label{eq:fl_prop_matrix_element}
\end{equation}
When all three components---the final state, the current, and the initial
state---are evolved consistently, then $\fL$ is unchanged.  However, if we miss
evolving a component, then we obtain a different result.  It is instructive to
illustrate this through a first-order analytical calculation.%
\footnote{An analogous calculation based on field redefinitions appears
in Ref.~\cite{Furnstahl:2001xq}.}

Let us look at the effects due to the evolution of individual components for a
general matrix element $\mbraket{\psi_f}{\widehat{O}}{\psi_i}$.  The evolved
initial state is given by
\begin{equation}
 \ket{\psi_i^\lambda} \equiv U \, \ket{\psi_i}
 = \ket{\psi_i} + \widetilde{U} \, \ket{\psi_i} \,,
\label{eq:psi_i_evolution}
\end{equation}
where $\widetilde{U}$ is the smooth part of the $U$-matrix defined in
Eq.~\eqref{eq:U_decomposition}.  Similarly, we can write down the expressions
for the evolved final state and the evolved operator as
\begin{equation}
 \bra{\psi_f^\lambda} \equiv \bra{\psi_f} \, U^\dag
 = \bra{\psi_f} - \bra{\psi_f}\,\widetilde{U}
\label{eq:psi_f_evolution}
\end{equation}
and
\begin{equation}
 \widehat{O}^\lambda
 \equiv U \, \widehat{O} \, U^\dag = \widehat{O}
 + \widetilde{U}\, \widehat{O} - \widehat{O} \, \widetilde{U}
 + \mathcal{O}(\widetilde{U}^2) \,.
\label{eq:O_evolution}
\end{equation}
We assume here that $\widetilde{U}$ is small compared to $I$ (which can always
be ensured by choosing the SRG $\lambda$ large enough) and therefore keep terms
only up to linear order in $\widetilde{U}$.  Using
Eqs.~\eqref{eq:psi_i_evolution}, \eqref{eq:psi_f_evolution},
and~\eqref{eq:O_evolution}, we get an expression for the evolved matrix element
in terms of the unevolved one and changes to individual components due to
evolution:
\begin{multline}
 \mbraket{\psi_f^\lambda}{\widehat{O}^\lambda}{\psi_i^\lambda}
 = \mbraket{\psi_f}{\widehat{O}}{\psi_i} - \underbrace{\mbraket{\psi_f}
  {\widetilde{U} \, \widehat{O}}{\psi_i}}_{\delta \bra{\psi_f}} \\
 \null + \underbrace{\mbraket{\psi_f}{\widetilde{U} \, \widehat{O}}{\psi_i}
 - \mbraket{\psi_f}{\widehat{O} \, \widetilde{U}}{\psi_i}}_{\delta\widehat{O}}
 + \underbrace{
  \mbraket{\psi_f} {\widehat{O} \, \widetilde{U}}{\psi_i}
 }_{\delta\ket{\psi_i}}
\end{multline}
\begin{equation}
 \Longrightarrow \mbraket{\psi_f^\lambda}{\widehat{O}^\lambda}{\psi_i^\lambda}
 = \mbraket{\psi_f}{\widehat{O}}{\psi_i} + \mathcal{O}(\widetilde{U}^2) \,.
\end{equation}
We see that the change due to evolution in the operator is equal and opposite
to the sum of changes due to the evolution of the initial and final states.
We also find that changes in each of the components are of the same order, and
that they mix;  this feature persists to higher order.  Therefore, if one misses
evolving an individual component, one will not reproduce the unevolved answer.

\subsection{Overview of numerical results}

For our analysis, we studied the effect of evolution of individual components
on $\fL$ for selected kinematics in the ranges $E^\prime = 10$--$100~\MeV$ and
$\mbf{q}^2 = 0.25$--$25~\fm^{-2}$, where $E^\prime$ is the energy of outgoing
nucleons and $\mbf{q}^2$ is the three-momentum transferred by the virtual
photon; both are taken in the center-of-mass frame of the outgoing nucleons.
This range was chosen to cover a variety of kinematics and motivated by the
set covered in Ref.~\cite{Yang:2013rza}.  We use the Argonne $v_{18}$ potential
(AV18)~\cite{Wiringa:1994wb} for our calculations.  It is one of the widely
used potentials for nuclear few-body reaction calculations, particularly those
involving large momentum transfers~\cite{Carlson:1997qn,Carlson:2014vla}.

How strong the evolution of individual components (or a subset thereof) affects
the result for $\fL$ depends on the kinematics.  One kinematic configuration of
particular interest is the so-called quasi-free ridge.  As discussed in
Sec.~\ref{subsec:disintegration_primer}, the four-momentum transferred by the
virtual photon in the center-of-mass frame is $(\omega, \mbf{q})$.  The
criterion for a configuration to lie on the quasi-free ridge is $\omega = 0$.
Physically, this means that the nucleons in the deuteron are on their mass
shell.  As shown in Ref.~\cite{Yang:2013rza}, at the quasi-free ridge the
energy of the outgoing nucleons ($E^\prime$) and the photon momentum transfer
are related by
\begin{equation}
 \Ep = \sqrt{M_d^2 + \mbf{q}^2} - 2 M \,,
 \label{eq:quasi_free_condition_exact}
\end{equation}
which reduces to
\begin{equation}
 E^\prime~\text{(in~$\MeV$)} \approx 10 \, \mbf{q}^2~\text{(in~$\fm^{-2}$)} \,.
\label{eq:quasi_free_condition}
\end{equation}
The quasi-free
condition in the center-of-mass frame is the same as the quasi-elastic
condition in the lab frame.  There, the quasi-elastic ridge is defined by $W^2
= m_p^2 \Rightarrow Q^2 = 2 \, \omega_\text{lab} \, m_p$, where $W$ is
the invariant mass.  On the quasi-elastic ridge, the so-called missing
momentum%
\footnote{The missing momentum is defined as the difference of
the measured proton momentum and the momentum transfer, $\mbf{p}_\text{miss}
\equiv \mbf{p}_\text{lab}^\text{proton} - \mbf{q}_\text{lab}$.}
vanishes, $p_\text{miss} = 0$.

\begin{figure}[htbp]
 \centering
 \includegraphics[width=0.98\columnwidth]
 {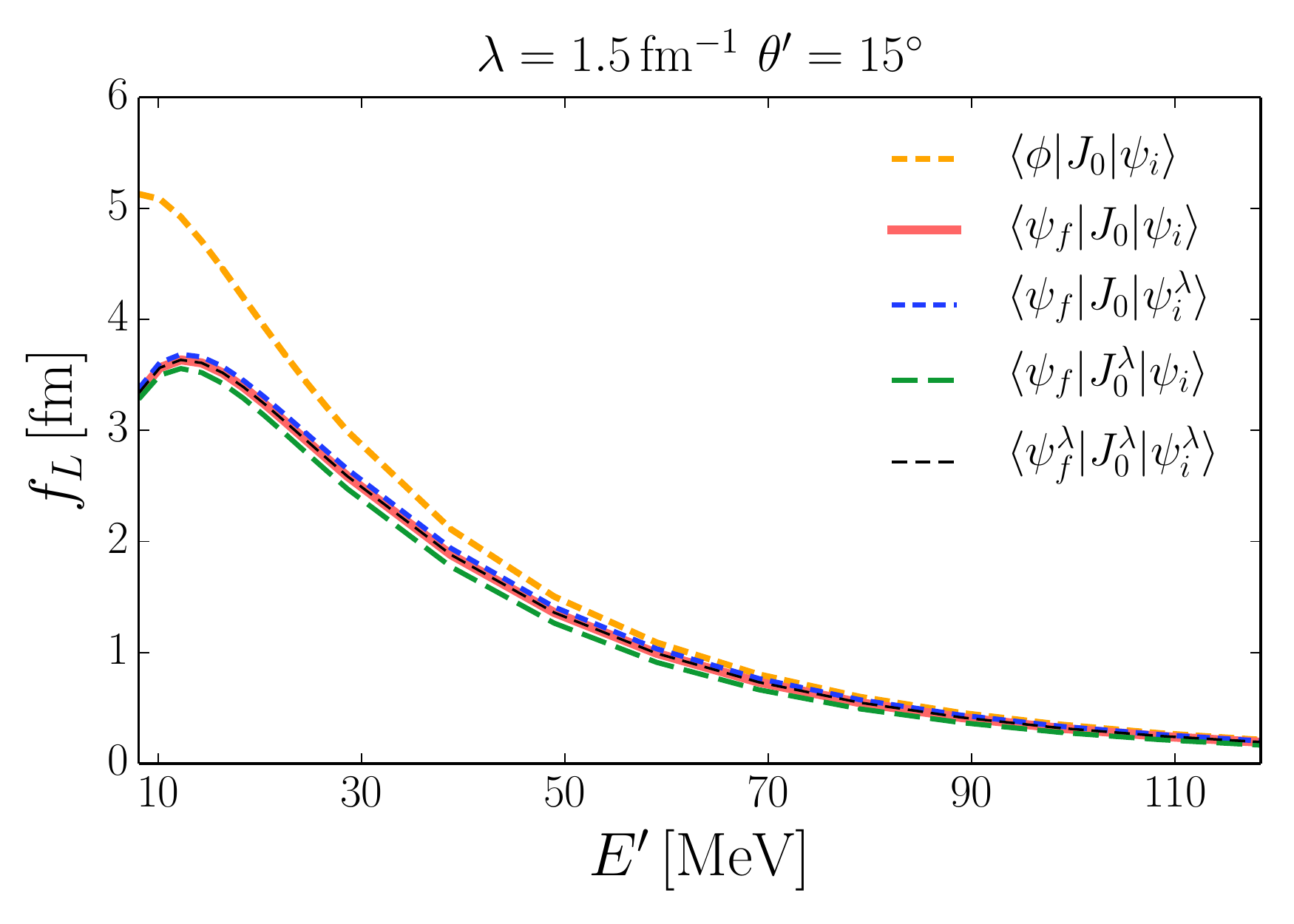}
 \caption{(color online)
   $\fL$ calculated at various points on the quasi-free ridge for
   $\thetacm = 15\degree$ for the AV18 potential.
   Legends indicate which component of the matrix element in
   Eq.~\eqref{eq:fl_prop_matrix_element} used to calculate $\fL$ is evolved.
   There are no appreciable evolution effects all
   along the quasi-free ridge.  The effect due to evolution of
   the final state is small as well and is not shown here to avoid clutter.
   $\fL$ calculated in the impulse
   approximation is also shown for comparison.}
 \label{fig:evolution_at_qfr}
\end{figure}

In Fig.~\ref{fig:evolution_at_qfr} we plot $\fL$ along the quasi-free ridge
both in the impulse approximation (IA) and with the final-state interactions
(FSI) included as a function of energy of the outgoing nucleons for a fixed
angle, $\thetacm = 15^{\degree}$ of the outgoing proton.
$\Ep$ and $\mbf{q}^2$ in Fig.~\ref{fig:evolution_at_qfr} are related by
Eq.~\eqref{eq:quasi_free_condition_exact}.  Comparing the solid curve labeled
$\mbraket{\psi_f}{J_0}{\psi_i}$ in the legend to the dashed curve (labeled
$\mbraket{\phi}{J_0}{\psi_i}$) we find that FSI effects are minimal for
configurations on the quasi-free ridge especially at large energies.

In an intuitive picture, this is because after the initial photon is absorbed,
both the nucleons in the deuteron are on their mass shell at the quasi-free
ridge, and therefore no FSI are needed to make the final-state particles real.
As we move away from the ridge, FSI become more important, as additional
energy-momentum transfer is required to put the neutron and the proton on shell
in the final state.  The difference between full $\fL$ and $\fL$ in IA at small
energies is also seen to hold for few-body nuclei \cite{Bacca:2014tla}.

Figure~\ref{fig:evolution_at_qfr} also shows $\fL$ calculated from evolving
only one of the components of the matrix element in
Eq.~\eqref{eq:fl_prop_matrix_element}.  We note that the effects of SRG
evolution of the individual components are minimal at the quasi-free ridge as
well.  The kinematics at the quasi-free ridge are such that only the long-range
(low-momentum) part of the deuteron wave function is probed, the FSI remains
small under evolution, and then unitarity implies minimal evolution of the
current.  As one moves away from the quasi-free ridge, the effects of evolution
of individual components become prominent.  Note that
$\mbraket{\psi_f}{J_0}{\psi_i} =
\mbraket{\psi_f^\lambda}{J_0^\lambda}{\psi_i^\lambda}$ and therefore the
unevolved vs.\ all-evolved $\fL$ overlap in Fig.~\ref{fig:evolution_at_qfr}.

\begin{figure}[htbp]
 \centering
 \includegraphics[width=0.85\columnwidth]{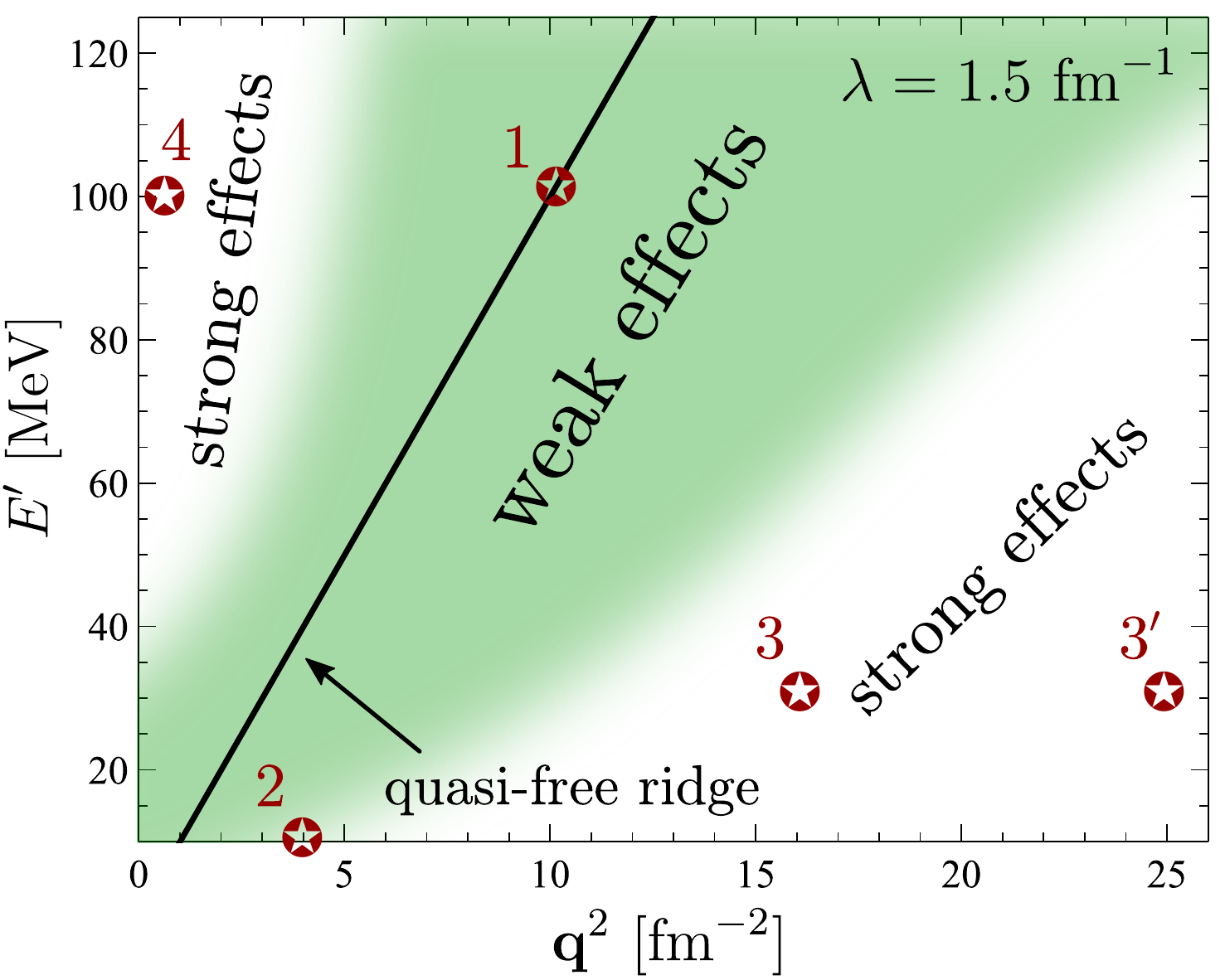}
 \caption{(color online) `Phase space' of kinematics for
 $\lambda = 1.5~\fm^{-1}$.  The effects of evolution
 get progressively prominent as one moves further away from the quasi-free
 ridge. The kinematics of the labeled points are considered in
 Sec.~\ref{subsec:illustrative_examples}.}
 \label{fig:More-1p5}
\end{figure}

Figure~\ref{fig:More-1p5} shows the `phase space' of kinematics for SRG
$\lambda = 1.5~\rm{fm^{-1}}$.  The quasi-free ridge is along the solid line
in Fig.~\ref{fig:More-1p5}.  In the shaded region the effects generated by the
evolution of individual components are weak (only a few percent
relative difference).
As one moves away from the quasi-free ridge, these differences get progressively
more prominent.
The terms `small' and `weak' in Fig.~\ref{fig:More-1p5} are used in a
qualitative
sense.  In the shaded region denoted by `weak effects', the effects of evolution
are not easily discernible on a typical $\fL$ versus $\thetacm$ plot, as
seen in Fig.~\ref{fig:100_10_quasi_free_fsi}, whereas in the region
labeled by `strong effects', the differences due to evolution are evident on
a plot (e.g., see Fig.~\ref{fig:30_16_fsi}).
The size of the shaded region in Fig.~\ref{fig:More-1p5} depends
on the SRG $\lambda$.  It is large for high $\lambda$'s and gets smaller as the
$\lambda$ is decreased (note that smaller SRG $\lambda$ means greater
evolution).  In the next subsection we look in detail at a few representative
kinematics, indicated by points in Fig.~\ref{fig:More-1p5}.

\subsection{Illustrative examples}
\label{subsec:illustrative_examples}

\subsubsection{At the quasi-free ridge}

\begin{figure}[htbp]
 \centering
 \includegraphics[width=0.98\columnwidth]{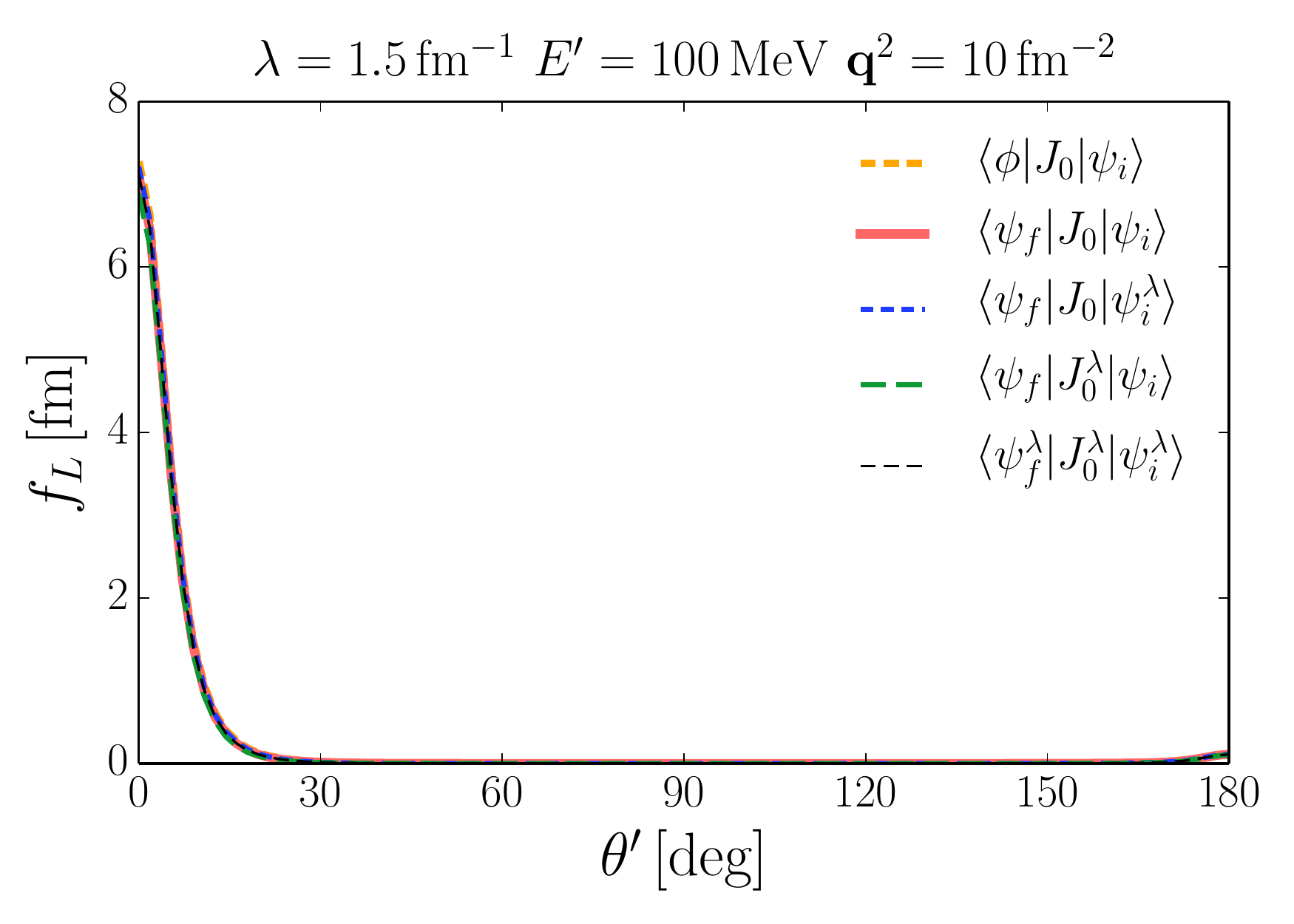}
 \caption{(color online)
   $\fL$ calculated for $E^\prime = 100~\MeV$ and $\mbf{q}^2 = 10~\fm^{-2}$
   (point ``1'' in Fig.~\ref{fig:More-1p5}) for the AV18 potential.
   Legends indicate which component of the matrix element in
   Eq.~\eqref{eq:fl_prop_matrix_element} used to calculate $\fL$ is evolved.
   $\thetacm$ is the angle of the outgoing proton in the center-of-mass frame.
   There are no discernible evolution effects for all
   angles.  The effect due to evolution
   of the final state is small as well and is not shown here to avoid clutter.
   $\fL$ calculated in the IA,
   $\mbraket{\phi}{J_0}{\psi_i}$, is also shown for comparison.
   }
 \label{fig:100_10_quasi_free_fsi}
\end{figure}

As a representative of quasi-free kinematics, we choose
$E^\prime = 100~\MeV$ and $\mbf{q}^2 = 10~\fm^{-2}$ and plot $\fL$ as a
function of angle in Fig.~\ref{fig:100_10_quasi_free_fsi}.  The effect of
including FSI is small for this configuration for all
angles.  Also, the effects due to evolution of the individual components are
too small to be discernible.  All this is consistent with the discussion in the
previous section.

\subsubsection{Near the quasi-free ridge}

\begin{figure}[htbp]
 \centering
 \includegraphics[width=0.98\columnwidth]{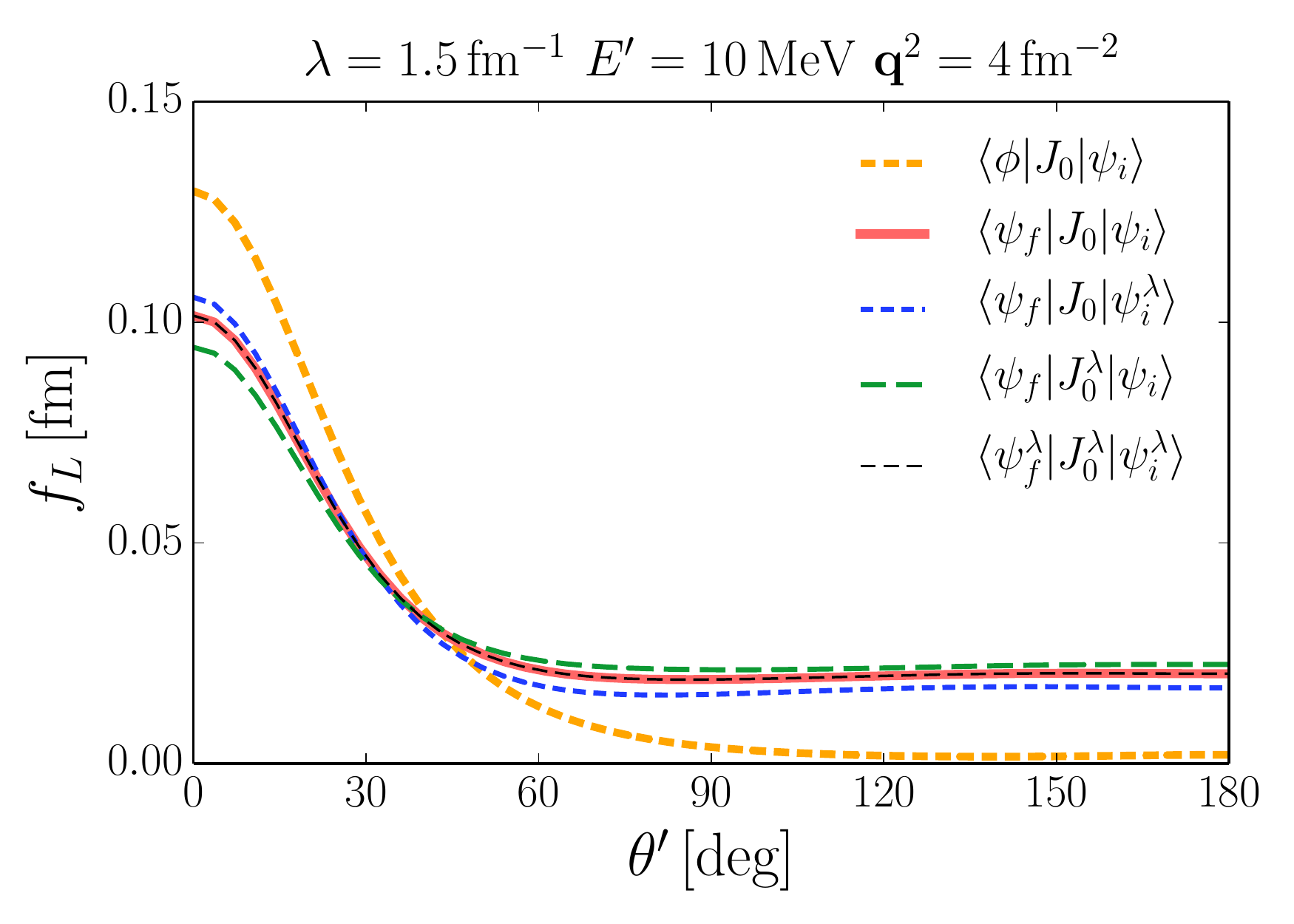}
 \caption{(color online) $\fL$ calculated for
 $E^\prime = 10~\MeV$ and $\mbfq^2 = 4~\fm^{-2}$
 (point ``2'' in Fig.~\ref{fig:More-1p5})
 for the AV18 potential.
 Legends indicate which component of the matrix element in
 Eq.~\eqref{eq:fl_prop_matrix_element} used to calculate $\fL$ is evolved.
 $\fL$ calculated in the IA, $\mbraket{\phi}{J_0}{\psi_i}$,
 is also shown for comparison.  The effects due to evolution of individual
 components on $\fL$ are discernible, but still small (compared to the FSI
 contribution).  The effect due to evolution of the final state is small as
 well and is not shown here to avoid clutter. }
 \label{fig:10_4_fsi}
\end{figure}

Next we look at the kinematics $E^\prime = 10~\MeV$ and $\mbf{q}^2 =
4~\fm^{-2}$, which is near the quasi-free ridge.  This is the point ``2'' in
Fig.~\ref{fig:More-1p5}.  As seen in Fig.~\ref{fig:10_4_fsi}, the different
curves for $\fL$ obtained from evolving different components start to diverge.
Figure~\ref{fig:10_4_fsi} also shows $\fL$ calculated in IA.  Comparing this to
the full $\fL$ including FSI, we see that the effects due to evolution are small
compared to the FSI contributions.  This smallness prevents us from making any
systematic observations about the effects due to evolution at this kinematics.
We thus move on to kinematics which show more prominent effects.

\subsubsection{Below the quasi-free ridge}

\begin{figure}[htbp]
 \centering
 \includegraphics[width=0.98\columnwidth]{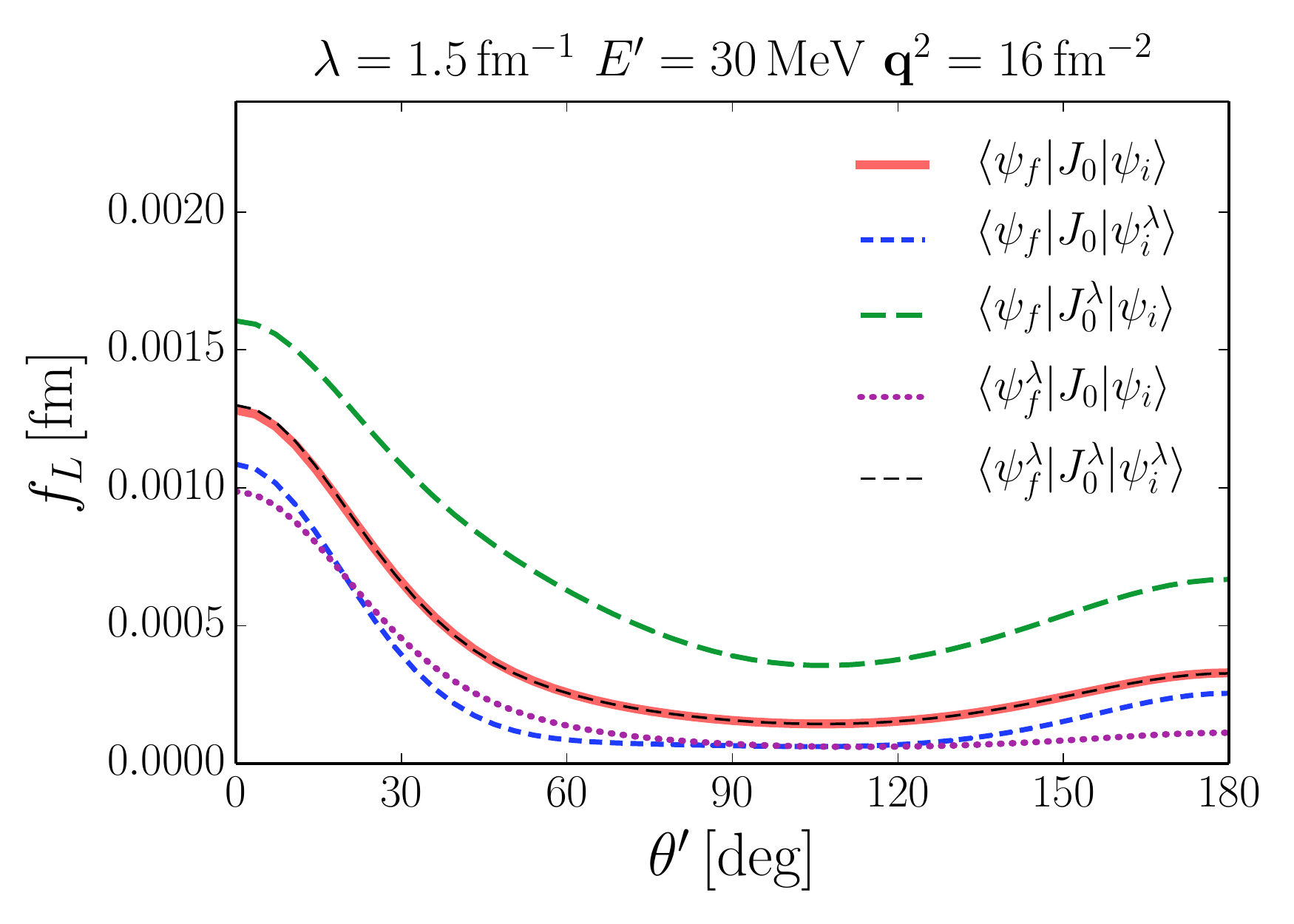}
 \caption{(color online) $\fL$ calculated for
 $E^\prime = 30~\MeV$ and $\mbfq^2 = 16~\fm^{-2}$
 (point ``3'' in Fig.~\ref{fig:More-1p5})
 for the AV18 potential.
 Legends indicate which component of the matrix element in
 Eq.~\eqref{eq:fl_prop_matrix_element} used to calculate $\fL$ is evolved.
 Prominent enhancement with evolution of the
 current only and suppression with evolution of the initial state and the final
 state only, respectively.}
 \label{fig:30_16_fsi}
\end{figure}

\begin{figure}[htbp]
 \centering
 \includegraphics[width=0.98\columnwidth]{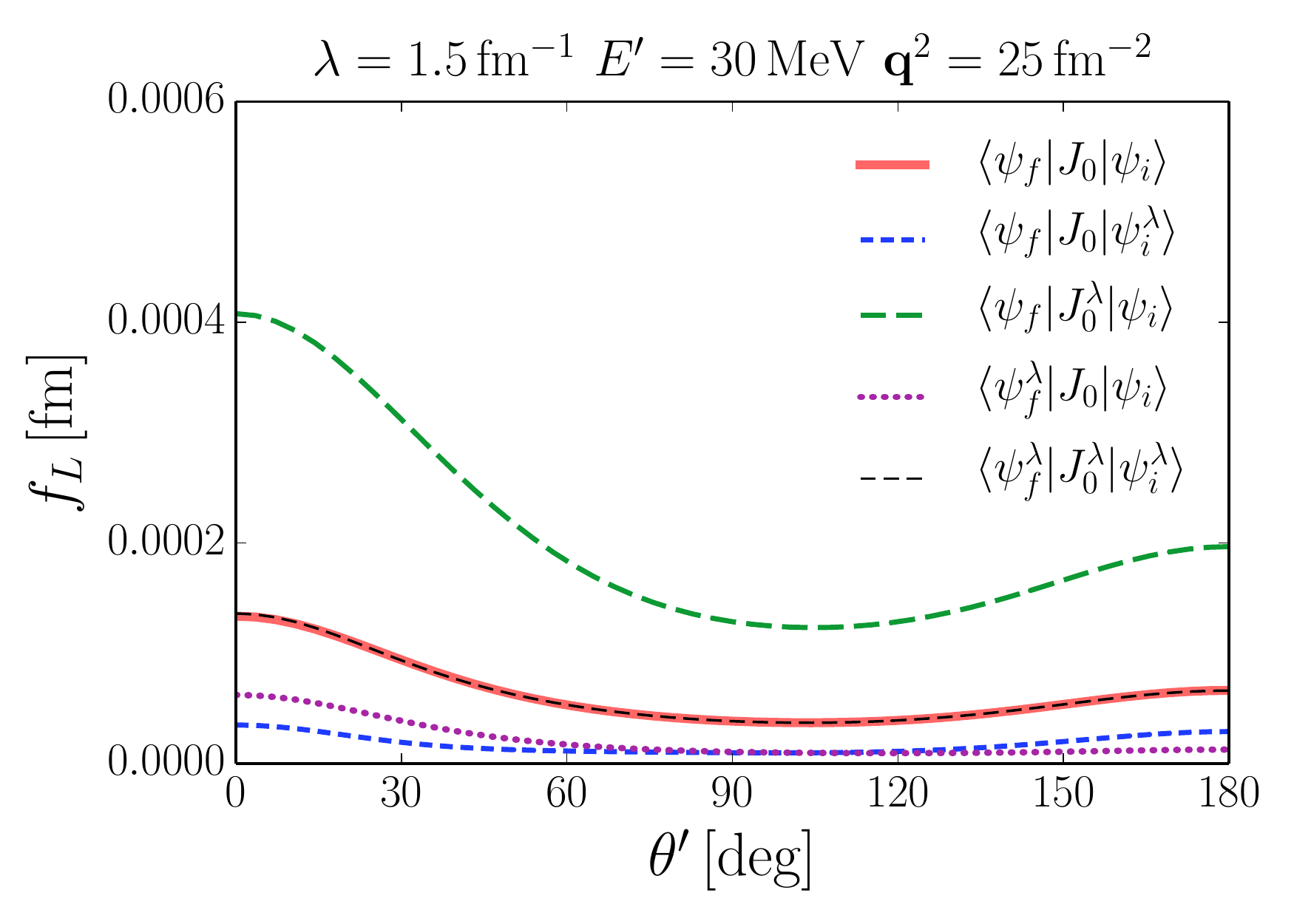}
 \caption{(color online) $\fL$ calculated for
 $E^\prime = 30~\MeV$ and $\mbfq^2 = 25~\fm^{-2}$
 (point ``3$^{\prime}$'' in Fig.~\ref{fig:More-1p5})
 for the AV18 potential.
 Legends indicate which component of the matrix element in
 Eq.~\eqref{eq:fl_prop_matrix_element} used to calculate $\fL$ is evolved.
 Prominent enhancement with evolution of the
 current only and suppression with evolution of the initial state and the final
 state only, respectively.}
 \label{fig:30_25_fsi}
\end{figure}

We next look in the region where $E^{\prime}~\text{(in~$\MeV$)}
\ll 10\,\mbfq^2~\text{(in~$\fm^{-2}$)}$, \ie, below the quasi-free ridge in
Fig.~\ref{fig:More-1p5}.  We look at two momentum transfers
$\mbfq^2 = 16~\fm^{-2}$ and $\mbfq^2 = 25~\fm^{-2}$ for $\Ep = 30~\MeV$,
which are points ``3'' and ``$3^\prime$'' in Fig.~\ref{fig:More-1p5}.
Figures~\ref{fig:30_16_fsi} and~\ref{fig:30_25_fsi} indicate the effects on $\fL$
from evolving individual components of the matrix elements.  It is noteworthy
that in both cases evolution of the current gives a prominent enhancement,
whereas evolution of the initial and final state gives a suppression.  When all
the components are evolved consistently, these changes combine and we recover
the unevolved answer for $\fL$.  This verifies the accurate implementation of
the equations derived in Sec.~\ref{sec:evolution_set_up}.


It is possible to qualitatively explain the behavior seen in
Figs.~\ref{fig:30_16_fsi} and~\ref{fig:30_25_fsi}.  As noted in
Eq.~\eqref{eq:overlap_IA_plus_FSI}, the overlap matrix element is given by the
sum of the IA part and the FSI part.  Below the quasi-free ridge these two
terms add constructively.  In this region, $\fL$ calculated in impulse
approximation is smaller than $\fL$ calculated by including the final-state
interactions.

\paragraph{Evolving the initial state}

\begin{figure}[htbp]
 \centering
 \includegraphics[width=0.8\columnwidth]{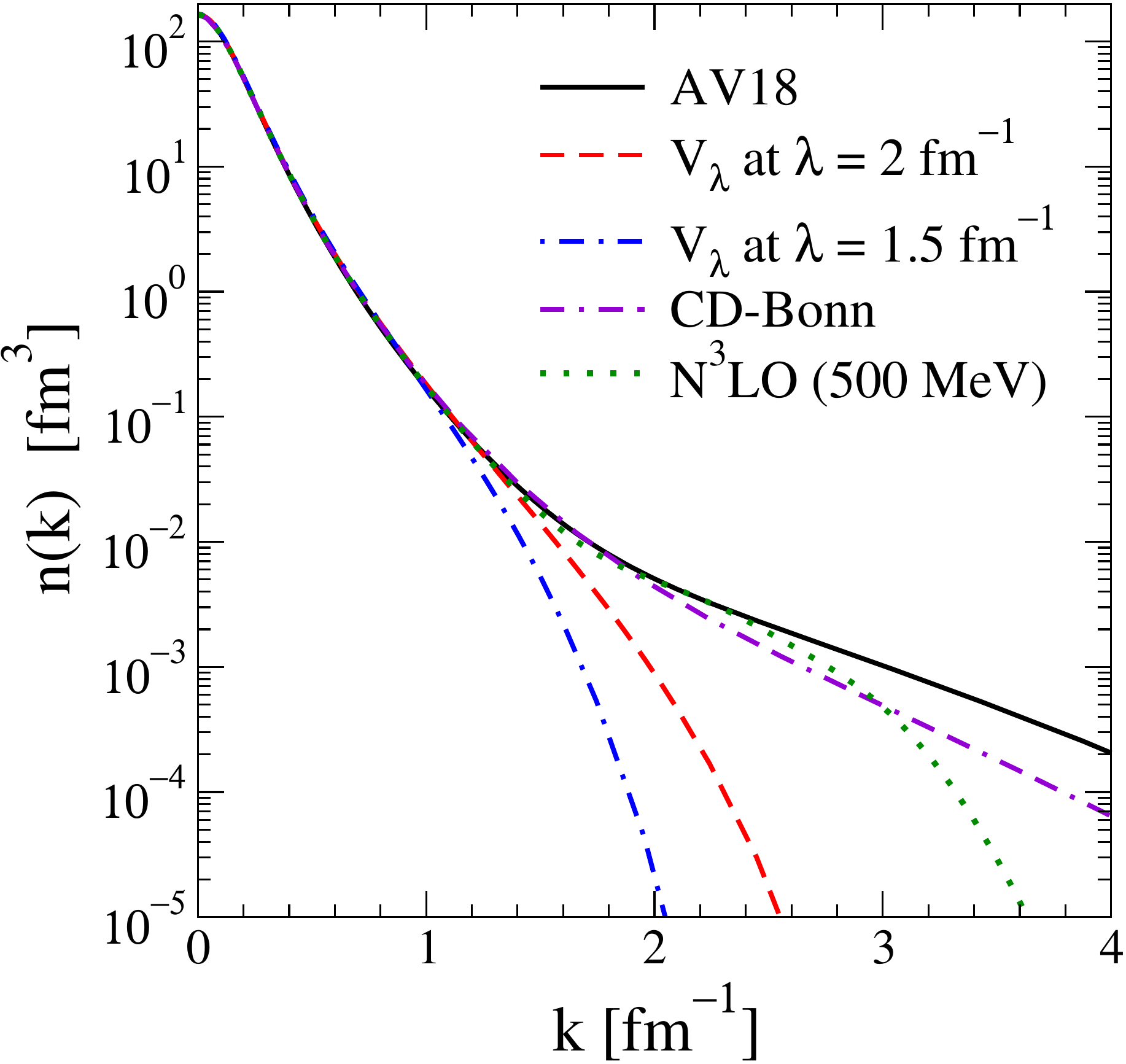}
 \caption{(color online) Momentum distribution
 for the deuteron for the AV18~\cite{Wiringa:1994wb},
 CD-Bonn~\cite{Machleidt:2000ge},
 and the Entem-Machleidt N$^3$LO chiral EFT~\cite{Entem:2003ft}
 potentials, and for the AV18 potential evolved to two SRG
 $\lambda$'s.}
 \label{fig:deut_md}
\end{figure}

Let us first consider the effect of evolving the initial state only.  We have
\begin{equation}
 \mbraket{\psi_f}{J_0}{\psi_i^\lambda}
 = \mbraket{\phi}{J_0}{\psi_i^\lambda}
 + \mbraket{\phi}{t^\dag \, \GreensFn^\dag \, J_0}{\psi_i^\lambda} \,.
\label{eq:overlap_IA_plus_FSI_evol_wf}
\end{equation}
As seen in Eq.~\eqref{eq:overlap_IA}, in the term
$\mbraket{\phi}{J_0}{\psi_i^\lambda}$ the deuteron wave function is
probed between $|\pp - q/2|$ and $\pp + q/2$.  These numbers are
$(1.2, 2.9)~\fm^{-1}$ and $(1.7, 3.4)~\fm^{-1}$ for $\Ep = 30~\MeV$,
$\mbfq^2 = 16~\fm^{-2}$ and $\Ep = 30~\MeV$, $\mbfq^2 = 25~\fm^{-2}$,
respectively.  The evolved deuteron wave function is significantly suppressed
at these high momenta.
This behavior is reflected in the deuteron momentum distribution plotted in
Fig.~\ref{fig:deut_md}.  The deuteron momentum distribution $n(k)$ is
proportional to the sum of the squares of $S$- and $D$- state deuteron
wave functions.
Thus, the first (IA) term in
Eq.~\eqref{eq:overlap_IA_plus_FSI_evol_wf} is much smaller than its
unevolved counterpart in Eq.~\eqref{eq:overlap_IA_plus_FSI}, for all angles.
We note that even though we only use the AV18 potential to
study changes due to evolution, these changes will be significant for other
potentials as well.

Evaluation of the second (FSI) term in
Eq.~\eqref{eq:overlap_IA_plus_FSI_evol_wf} involves an integral over all
momenta, as indicated in Eq.~\eqref{eq:phi_t_g0_J0_minus}.  We find that
$|\mbraket{\phi}{t^\dag \, \GreensFn^\dag \, J_0}{\psi_i^\lambda}| <
|\mbraket{\phi}{t^\dag \, \GreensFn^\dag \, J_0}{\psi_i}|$.  As mentioned
before, because the terms $\mbraket{\phi}{J_0}{\psi_i}$ and
$\mbraket{\phi}{t^\dag \, \GreensFn^\dag \, J_0}{\psi_i}$ add constructively
below the quasi-free ridge and because the magnitude of both these terms
decreases upon evolving the wave function, we have
\begin{equation}
 |\mbraket{\psi_f}{J_0}{\psi_i^\lambda}| < |\mbraket{\psi_f}{J_0}{\psi_i}|
\label{eq:evolv_wf_below_qfr} \,.
\end{equation}
The above relation holds for most combinations of $\mJd$ and $\msf$.  For those
$\mJd$ and $\msf$ for which Eq.~\eqref{eq:evolv_wf_below_qfr} does not hold,
the absolute value of the matrix element is much smaller than for those for
which the Eq.~\eqref{eq:evolv_wf_below_qfr} \emph{does} hold, and therefore we
have $\fL$ calculated from $\mbraket{\psi_f}{J_0}{\psi_i^\lambda}$ smaller than the
$\fL$ calculated from $\mbraket{\psi_f}{J_0}{\psi_i}$, as seen in
Figs.~\ref{fig:30_16_fsi} and~\ref{fig:30_25_fsi}.

\paragraph{Evolving the final state}

As indicated in Eq.~\eqref{eq:psi_f_lam_def}, evolving the final state entails
the evolution of the $t$-matrix.  The overlap matrix element therefore is
\begin{equation}
 \mbraket{\psi_f^\lambda}{J_0}{\psi_i}
 = \mbraket{\phi}{J_0}{\psi_i}
 + \mbraket{\phi}{t^\dag_\lambda \, \GreensFn^\dag \, J_0}{\psi_i} \,.
\label{eq:evol_final_state_overlap}
\end{equation}
The IA term is the same as in the unevolved case.  The SRG evolution leaves the
on-shell part of the $t$-matrix---which is directly related to
observables---invariant.  The magnitude of the relevant off-shell $t$-matrix
elements decreases
on evolution, though.  As a result we have
\begin{equation}
|\mbraket{\psi_f^\lambda}{J_0}{\psi_i}| < |\mbraket{\psi_f}{J_0}{\psi_i}| \,.
\end{equation}
This is reflected in $\fL$ as calculated from the evolved final state, and
seen in Figs.~\ref{fig:30_16_fsi} and~\ref{fig:30_25_fsi}.

The effect of evolution of the initial state and the final state is to suppress
$\fL$.  When all the three components are evolved, we reproduce the unevolved
answer as indicated in Fig.~\ref{fig:30_16_fsi} and~\ref{fig:30_25_fsi}.  It is
therefore required that we find a huge enhancement when just the current
is evolved.

The kinematics $\Ep = 30~\MeV$, $\mbfq^2 = 25~\fm^{-2}$ is further away from
the quasi-free ridge than $\Ep = 30~\MeV$, $\mbfq^2 = 16~\fm^{-2}$.
The evolution effects discussed above get
progressively more prominent the further away one is from the quasifree
ridge.  This can be verified by
comparing the effects due to evolution of individual components in
Figs.~\ref{fig:30_16_fsi} and~\ref{fig:30_25_fsi}.

As remarked earlier, away from the quasi-free ridge the FSI
become important.  Nonetheless, it is still instructive to look at $\fL$
calculated in the IA at these kinematics.
%
\begin{figure}[htbp]
 \centering
 \includegraphics[width=0.98\columnwidth]{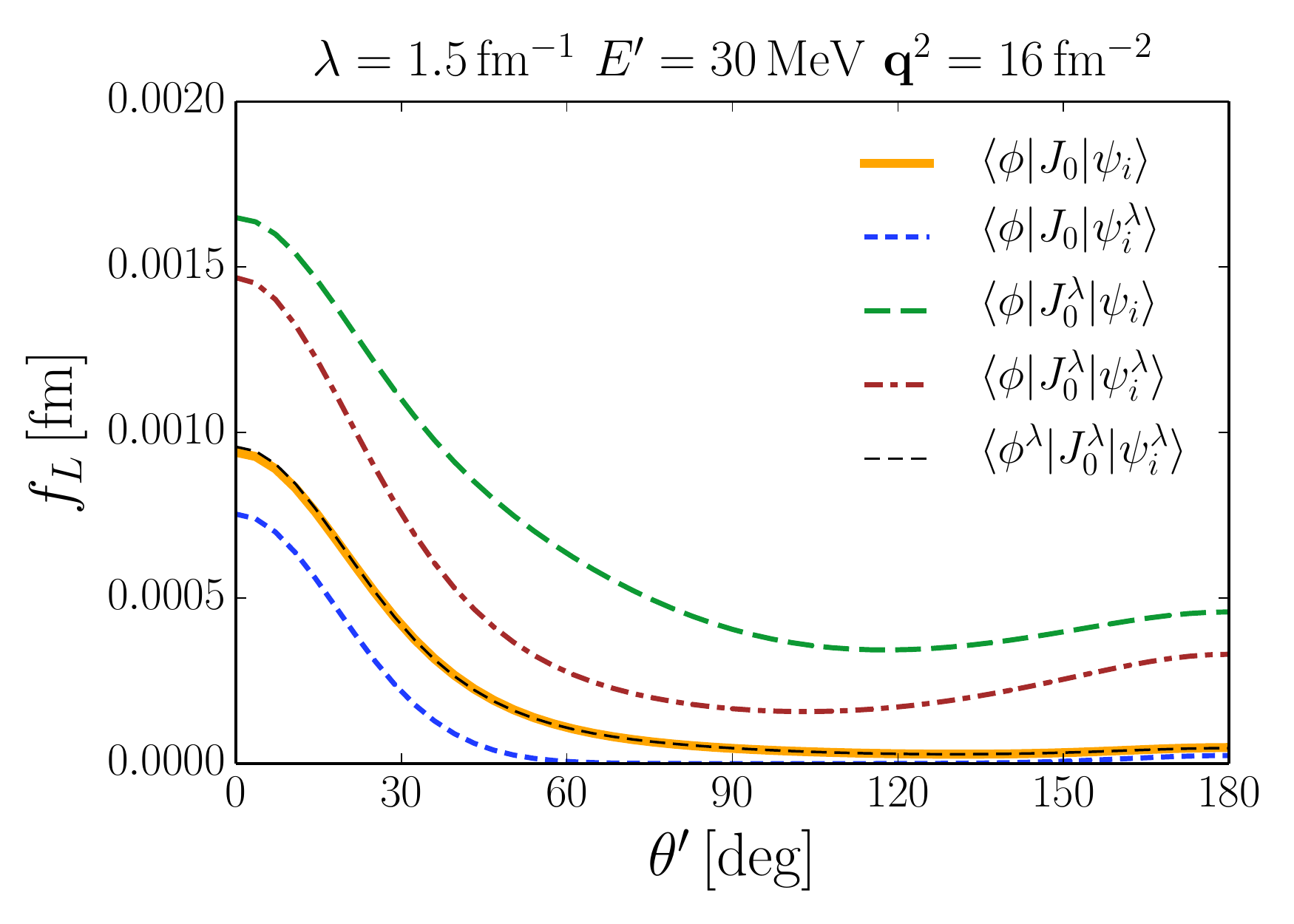}
 \caption{(color online) $\fL$ in IA
 $(\bra{\psi_f} \equiv \bra{\phi})$ calculated for
 $E^\prime = 30~\MeV$ and $\mbfq^2 = 16~\fm^{-2}$
 for the AV18 potential.
 Legends indicate which component of the matrix element in
 Eq.~\eqref{eq:fl_prop_matrix_element} used to calculate $\fL$ are evolved. }
 \label{fig:30_16_ia}
\end{figure}
\begin{figure}[htbp]
 \centering
 \includegraphics[width=0.98\columnwidth]{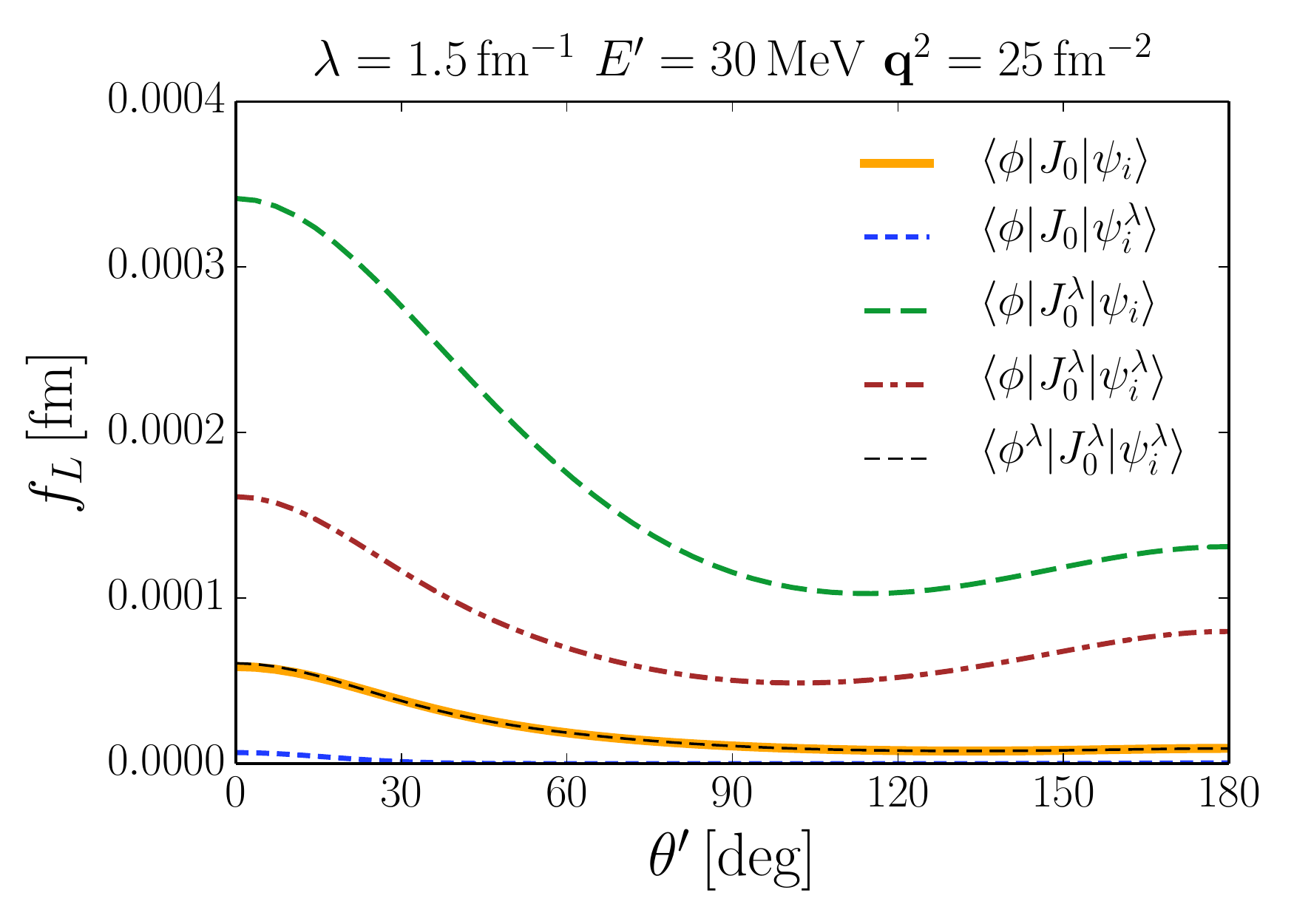}
 \caption{(color online) $\fL$ in IA
 $(\bra{\psi_f} \equiv \bra{\phi})$ calculated for
 $E^\prime = 30~\MeV$ and $\mbfq^2 = 25~\fm^{-2}$
 for the AV18 potential.
 Legends indicate which component of the matrix element in
 Eq.~\eqref{eq:fl_prop_matrix_element} used to calculate $\fL$ are evolved.}
 \label{fig:30_25_ia}
\end{figure}
%
Note that the (unevolved) $\fL$ calculated in the IA, shown
in Figs.~\ref{fig:30_16_ia} and~\ref{fig:30_25_ia}, is smaller than the full
$\fL$ that takes into account the final state interactions (\cf~the
corresponding curves in Figs.~\ref{fig:30_16_fsi} and~\ref{fig:30_25_fsi}).
This is consistent with the claim made earlier that below the quasi-free ridge
the two terms in Eq.~\eqref{eq:overlap_IA_plus_FSI} add constructively.

The results in Figs.~\ref{fig:30_16_ia} and~\ref{fig:30_25_ia} can again be
qualitatively explained based on our discussion above.  The evolution of the
deuteron wave function leads to suppression as the evolved wave function does
not have strength at high momentum.  The evolved current thus leads to
enhancement.  Evolution of both the current and the initial state decreases
$\fL$ from just the evolved current value, but it is not until we evolve all
three components---final state, current, and the initial state---that we
recover the unevolved answer.

\begin{figure}[htbp]
  \centering
 \includegraphics[width=0.98\columnwidth]{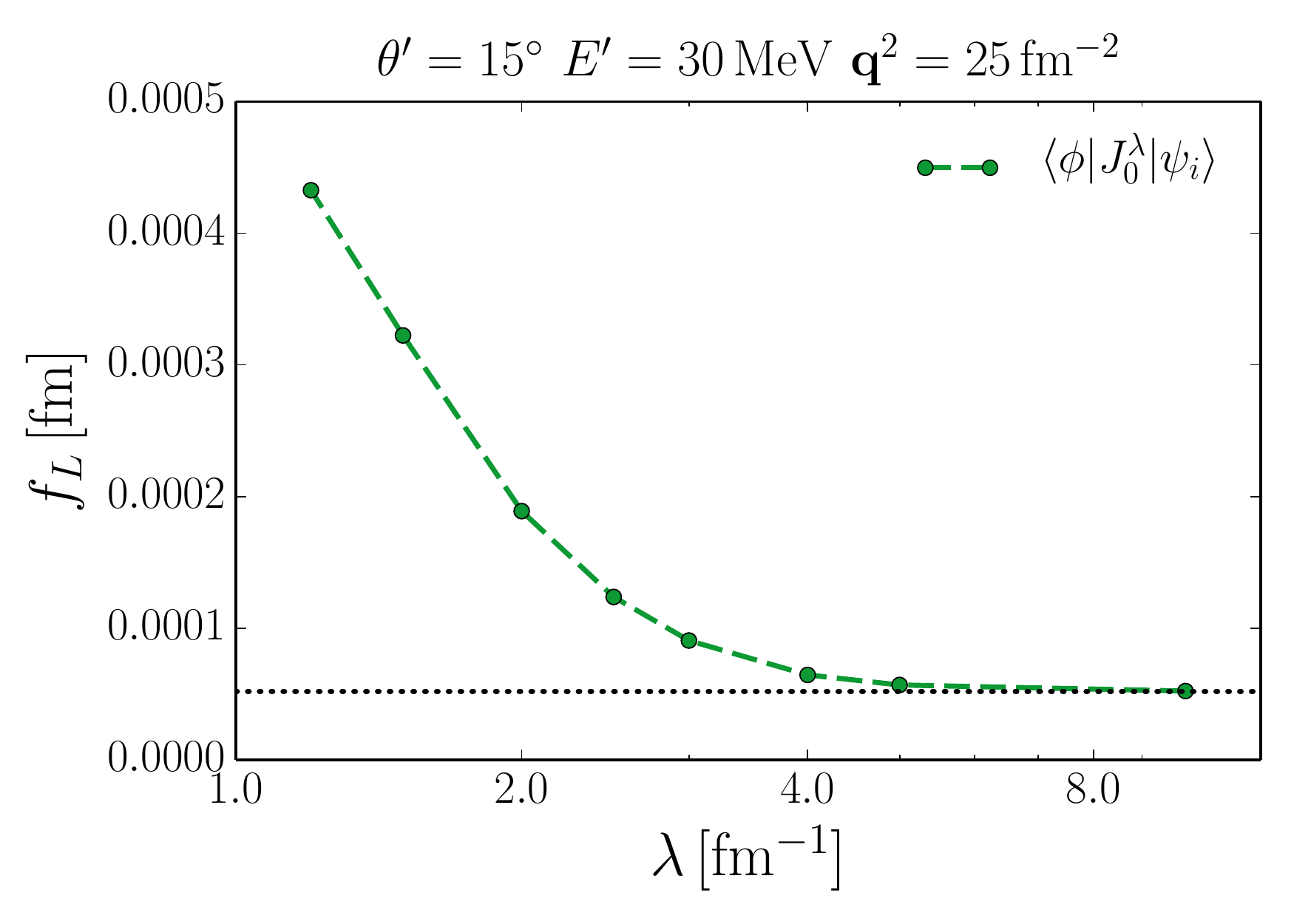}
 \caption{(color online) $\fL$ in IA
 calculated at $\thetacm = 15\degree$ for
 $E^\prime = 30~\MeV$ and $\mbfq^2 = 25~\fm^{-2}$
 for the AV18 potential when the current operator in
 Eq.~\eqref{eq:fl_prop_matrix_element} used to calculate $\fL$ is evolved to
 various SRG $\lambda$'s.  The horizontal dotted line is the unevolved
 answer.}
 \label{fig:J0_evolution_vs_lambda}
\end{figure}

As expected, the effect due to evolution increases with further evolution.
This is illustrated in Fig.~\ref{fig:J0_evolution_vs_lambda}, where we
investigate the effects of the current-operator evolution on $\fL$ as a
function of the SRG $\lambda$.  To isolate the effect of operator evolution, we
only look at $\fL$ calculated in IA at a specific angle in
Fig.~\ref{fig:J0_evolution_vs_lambda}.

\subsubsection{Above the quasi-free ridge}

\begin{figure}[htbp]
 \centering
 \includegraphics[width=0.98\columnwidth]{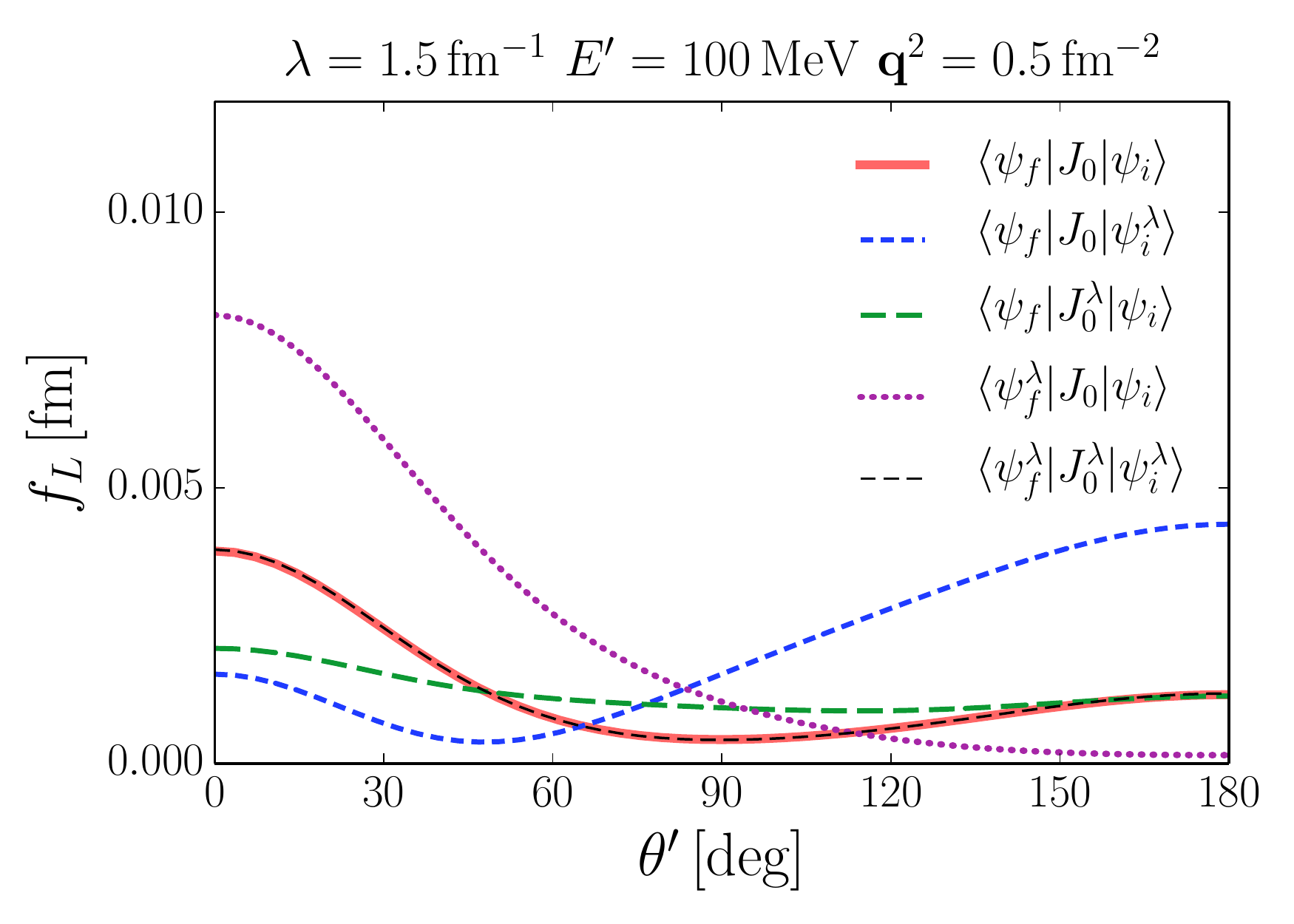}
 \caption{(color online) $\fL$ calculated for
 $E^\prime = 100~\MeV$ and $\mbfq^2 = 0.5~\fm^{-2}$
 (point ``4'' in Fig.~\ref{fig:More-1p5})
 for the AV18 potential.
 Legends indicate which component of the matrix element in
 Eq.~\eqref{eq:fl_prop_matrix_element} used to calculate $\fL$ is evolved.
 Opposite effects from the evolution of the
 initial state and the final state.}
 \label{fig:100_0p5_fsi}
\end{figure}

Finally, we look at an example from above the quasi-free ridge.
Figure~\ref{fig:100_0p5_fsi} shows the effect of evolution of individual
components on $\fL$ for $\Ep = 100~\MeV$ and $\mbfq^2 = 0.5~\fm^{-2}$, which is
point ``4'' in Fig.~\ref{fig:More-1p5}.  The effects of evolution in this case
are qualitatively different from those found below the quasi-free ridge.  For
instance, we see a peculiar suppression in $\fL$ calculated from the evolved
deuteron wave function at small angles, but an enhancement at large angles.  An
opposite behavior is observed for the final state.  It is again possible to
qualitatively explain these findings.

\paragraph{Evolving the initial state}

\begin{figure}[htbp]
 \centering
 \includegraphics[width=0.98\columnwidth]{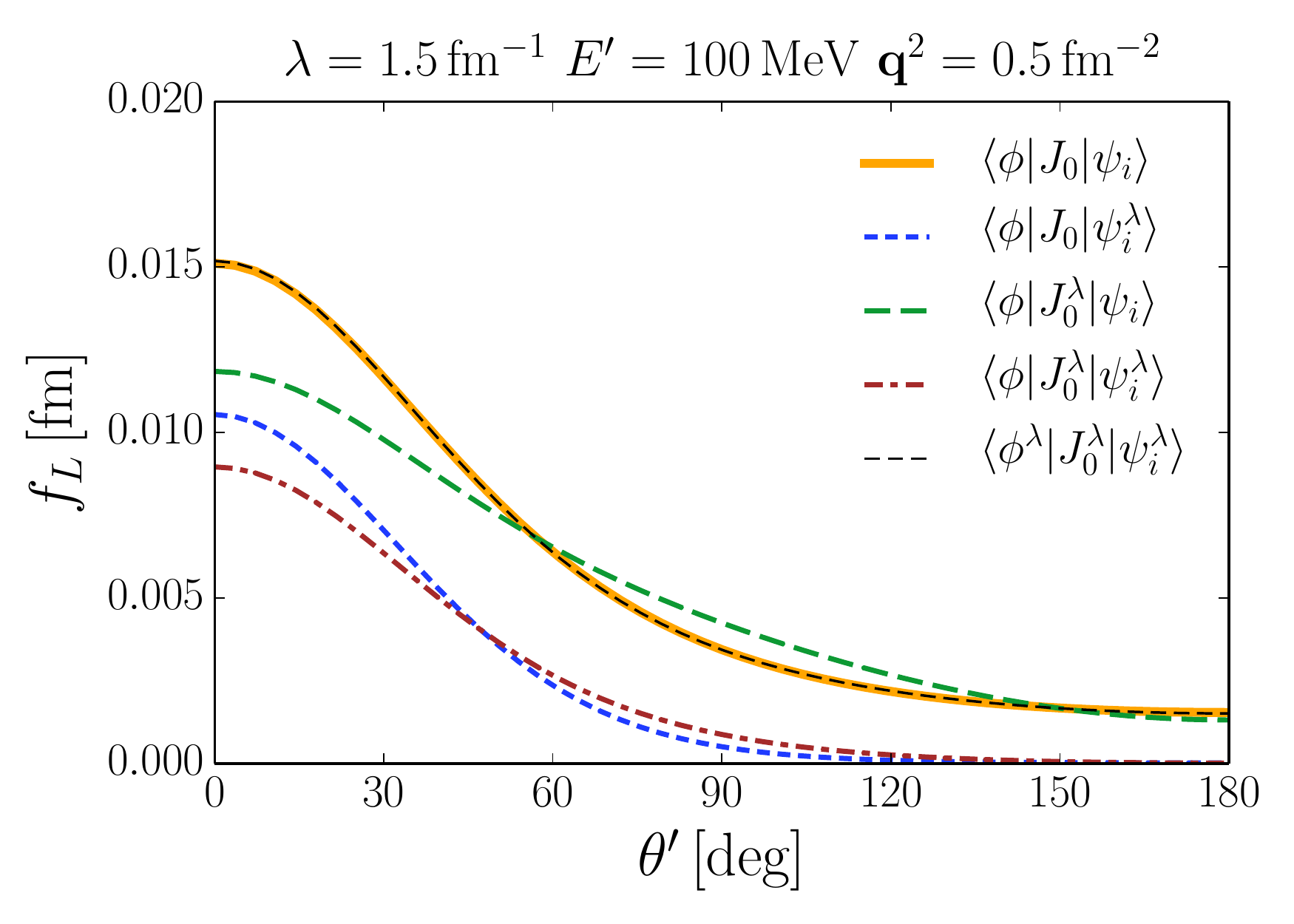}
 \caption{(color online) $\fL$ in IA
 $(\bra{\psi_f} \equiv \bra{\phi})$ calculated for
 $E^\prime = 100~\MeV$ and $\mbfq^2 = 0.5~\fm^{-2}$
 for the AV18 potential.
 Legends indicate which component of the matrix element in
 Eq.~\eqref{eq:fl_prop_matrix_element} used to calculate $\fL$ are evolved.}
 \label{fig:100_0p5_ia}
\end{figure}

Above the quasi-free ridge, the IA and FSI terms in
Eq.~\eqref{eq:overlap_IA_plus_FSI} add destructively.  This can be seen by
comparing the unevolved $\fL$ curves in Figs.~\ref{fig:100_0p5_fsi}
and~\ref{fig:100_0p5_ia}.  Including the FSI brings down
the value of $\fL$ when one is above the quasi-free ridge.

At small angles, the magnitude of the IA term in
Eq.~\eqref{eq:overlap_IA_plus_FSI} is larger than that of the FSI term.  The
deuteron wave function for this kinematics is probed between $1.2$ and
$1.9~\fm^{-1}$.  With the wave-function evolution, the magnitude of the IA
term in Eq.~\eqref{eq:overlap_IA_plus_FSI_evol_wf} decreases, whereas the
magnitude of the FSI term in that equation slightly increases compared to its
unevolved counterpart.  Still, at small angles, we have
$|\mbraket{\phi}{J_0}{\psi_i^\lambda}| > |\mbraket{\phi}{t^\dag \,
\GreensFn^\dag \, J_0}{\psi_i^\lambda}|$, which leads to
\begin{equation}
 |\mbraket{\psi_f}{J_0}{\psi_i^\lambda}| < |\mbraket{\psi_f}{J_0}{\psi_i}| \,,
 \label{eq:psi_i_evolution_small_angles_above_qfr}
\end{equation}
and thus to the suppression of $\fL$ at small angles observed in
Fig.~\ref{fig:100_0p5_fsi}.

At large angles, the magnitude of the IA term in
Eq.~\eqref{eq:overlap_IA_plus_FSI} is smaller than that of the FSI term.
With the wave-function evolution, the magnitude of IA term decreases
substantially (large momenta in the deuteron wave function are probed at large
angles, \cf~Eq.~\eqref{eq:overlap_IA}), whereas the FSI term in
Eq.~\eqref{eq:overlap_IA_plus_FSI} remains almost the same.  This results in
increasing the difference between the two terms in
Eq.~\eqref{eq:overlap_IA_plus_FSI} as the SRG $\lambda$ is decreased.  As
mentioned before, above the quasi-free ridge, the IA and FSI terms
in Eq.~\eqref{eq:overlap_IA_plus_FSI} add destructively and we therefore end up
with $|\mbraket{\psi_f}{J_0}{\psi_i^\lambda}| >
|\mbraket{\psi_f}{J_0}{\psi_i}|$, leading to the observed enhancement at large
angles upon evolution of the wave function (see Fig.~\ref{fig:100_0p5_fsi}).

\paragraph{Evolving the final state}

The expression to consider is Eq.~\eqref{eq:evol_final_state_overlap}.  With
the evolution of the $t$-matrix, the magnitude of the term
$\mbraket{\phi}{t^\dag_\lambda \, \GreensFn^\dag \, J_0}{\psi_i}$
decreases, and because of the opposite relative signs of the two terms in
Eq.~\eqref{eq:evol_final_state_overlap}---and
because at small angles the magnitude of the IA term is larger than the
FSI term---the net
effect is $|\mbraket{\psi_f^\lambda}{J_0}{\psi_i}| >
|\mbraket{\psi_f}{J_0}{\psi_i}|$.  This leads to an enhancement of $\fL$ with
evolved final state at small angles, as seen in Fig.~\ref{fig:100_0p5_fsi}.

At large angles the magnitude of the IA term in
Eq.~\eqref{eq:evol_final_state_overlap} is smaller than that of the FSI term.
With the evolution of the $t$-matrix, the magnitude of the FSI term decreases
and the difference between the IA and the FSI terms decreases as well.  This
leads to the observed overall suppression in $\fL$ at large angles due to the
evolution of the final state seen in Fig.~\ref{fig:100_0p5_fsi}.  For those few
($\msf$, $\mJd$) combinations for which the above general observations do not
hold, the value of individual components is too small to make any qualitative
difference.

Figure~\ref{fig:100_0p5_ia} shows the effect of evolution of individual
components on $\fL$ calculated in the IA for the kinematics
under consideration.  Again the evolved deuteron wave function does not have
strength at high momenta and therefore $\fL$ calculated from
$\mbraket{\phi}{J_0}{\psi_i ^\lambda}$ has a lower value than its unevolved
counterpart.

Unitary evolution means that the effect of the evolved current is always
such that it compensates the effect due to the evolution of the initial and
final states.  In future work we will examine more directly the behavior of the
current as it evolves to better understand how to carry over the results
observed here to other reactions.

\section{Summary and outlook}
\label{sec:conclusion}

Nuclear properties such as momentum distributions are extracted from experiment
by invoking the factorization of structure, which includes descriptions of
initial and final states, and reaction, which includes the description of the
probe components.
The factorization between reaction and structure depends on the scale and
scheme chosen for doing calculations.  Unlike in high-energy QCD, this scale
and scheme dependence of factorization is often not taken into account in
low-energy nuclear physics calculations, but is potentially critical for
interpreting experiment.
In our work we investigated this issue by looking at the simplest knockout
reaction: deuteron electrodisintegration.  We used SRG transformations to test
the sensitivity of the longitudinal structure function~$\fL$ to evolution of its
individual components: initial state, final state, and the current.

We find that the effects of evolution depend on kinematics,
but in a \emph{systematic}
way.  Evolution effects are negligible at the quasi-free ridge,
indicating that the
scale dependence of individual components is minimal there.  This is consistent
with the quasi-free ridge mainly probing the long-range part of the
wave function, which is largely invariant under SRG evolution.  This is also the
region where contributions from FSI to $\fL$ are minimal.
The effects get progressively more pronounced the further one moves away from
the quasifree ridge.  The nature of these changes depends on whether one
is above or below the quasifree ridge in the ‘“phase-space’” plot (Fig. 3).
As indicated in Sec. IIIC, these changes can also be explained qualitatively
by looking at the overlap matrix elements.  This allows us to predict the
effects due to evolution depending on kinematics.

Our results demonstrate that scale dependence needs to be taken into account
for low-energy nuclear calculations.  While we showed this explicitly only for
the case of the longitudinal structure function in deuteron disintegration,
we expect the results should qualitatively
carry over for other knock-out reactions as well.
An area of active investigation is the extension of the formalism presented here
to hard scattering processes.

SRG transformations are routinely used in nuclear structure calculations because
they lead to accelerated convergence for observables like binding energies.  We
demonstrated that SRG transformations can be used for nuclear knock-out
reactions as well as long as the operator involved is also consistently evolved.
The evolved operator appears to be complicated compared to the unevolved one,
but the factorization might be cleaner with the evolved operator if one can
exploit an operator product expansion~\cite{Anderson:2010aq,Bogner:2012zm}.
We plan to explore this in our future work.

We also plan to use pionless EFT as a framework to quantitatively study the
effects of operator evolution.  It should be a good starting point to
understand in detail how a one-body operator develops strength in two- and
higher-body sectors upon evolution.  This can give insight on the issue of power
counting of operator evolution.  Pionless EFT has been employed previously to
study deuteron electrodisintegration in Ref.~\cite{Christlmeier:2008ye}, where
it was used to resolve a discrepancy between theory and experiment.

Extending our work to many-body nuclei requires inclusion of 3N forces and 3N
currents.  Consistent evolution in that case would entail evolution in both two
and three-body sectors.  However, SRG transformations have proven to be
technically feasible for evolving three-body
forces~\cite{Jurgenson:2009qs,Jurgenson:2010wy,Hebeler:2012pr,Wendt:2013bla}.
Thus, extending our calculations to many-body nuclei would be computationally
intensive, but is feasible in the existing framework.  Including the effects
of FSI is challenging for many-body systems and has been
possible only recently for light nuclei \cite{Bacca:2014tla, Lovato:2015qka}.
It would be interesting to investigate if the scale and scheme dependence of
factorization allows us to choose a scale where the FSI effects are minimal.

\begin{acknowledgments}
We are grateful to S.~Jeschonnek, A.~Schwenk, and H.~Hergert for useful
discussion.  We would also like to thank H.~Arenh\"{o}vel and C.-J.~Yang for 
patiently answering our questions about the deuteron electrodisintegration 
formalism, as well as for sharing their data with us to compare results.  This 
work was supported in part by the National Science Foundation under Grant 
No.~PHY--1306250, the NUCLEI SciDAC Collaboration under DOE Grant DE-SC0008533 
and by the ERC Grant No.~307986 STRONGINT.
\end{acknowledgments}

\appendix

\section{Expressions from evolution}
\label{appendix:evol_eqns}

Here we document the expressions used in Sec.~\ref{sec:evolution_set_up}.
As seen in Eq.~\eqref{eq:B_split_up}, in order to evaluate the term $\la \phi |
\, J_0^\lambda | \psi_i^\lambda \ra$ we split it into four terms: $B_1$,
$B_2$, $B_3$, and $B_4$.  $B_4$ is obtained from Eq.~\eqref{eq:overlap_IA} by
using the evolved deuteron wave function instead of the unevolved one.  The
expressions for the terms $B_3$, $B_2$, and $B_1$ are as follows:
\begin{widetext}
\begin{multline}
 B_3 \equiv \la \phi | \, J_0 \, \widetilde{U}^\dag | \psi_i^\lambda \ra
 = 2 \, \sqrt{\frac{2}{\pi}} \,
 \sum_{T_1 = 0,1} \big(G_E ^p + (-1)^{T_1} \, G_E^n\big)
 \sum_{L_1 = 0}^{L_{\rm max}} \big(1 + (-1)^{T_1} (-1)^{L_1}\big) \,
  Y_{L_1 , \mJd - \msf} (\thetacm, \phicm) \\
 \null \times \sum_{J_1 = |L_1 - 1|}^{L+1}
  \CG{L_1}{\mJd - \msf}{S=1}{\msf}{J_1}{\mJd} \,
 \sum_{\widetilde{m}_s = -1}^1
  \CG*{J_1}{\mJd}{L_1}{\mJd - \widetilde{m}_s}{S=1}{\widetilde{m}_s} \\
 \null \times \sum_{L_2 = 0}^{L_{\rm max}}
  \CG{L_2}{\mJd-\widetilde{m}_s}{S=1}{\widetilde{m}_s}{J=1}{\mJd}
 \sum_{L_d = 0, 2}
  \int \! \dd k_3 \, \psi_{L_d}^\lambda(k_3) \, k_3^2 \int \! \dcostheta \,
  P_{L_1}^{\mJd-\widetilde{m}_s}(\cos \theta) \\
  \null \times
  P_{L_2}^{\mJd-\widetilde{m}_s}\!\big(\cos\thetacprime(\pp, \theta)\big) \,
  \widetilde{U}\!\left(
   k_3, \sqrt{{\pp}^2 - \pp q \cos\theta + q^2/4}, L_d, L_2, J=1, S=1, T=0
  \right) \,,
\end{multline}
\begin{multline}
 B_2 \equiv \la \phi | \,\widetilde{U}\, J_0 \, | \psi_i^\lambda \ra
 = 2 \, \sqrt{\frac{2}{\pi}} \,
 \sum_{T_1 = 0,1} \big(G_E ^p + (-1)^{T_1} \, G_E^n\big)
 \sum_{L_1 = 0}^{L_{\rm max}} \big(1 + (-1)^{T_1} (-1)^{L_1}\big) \,
  Y_{L_1 , \mJd - \msf} (\thetacm, \phicm) \\
 \null \times \sum_{J_1 = |L_1 - 1|}^{L+1}
  \CG{L_1}{\mJd - \msf}{S=1}{\msf}{J_1}{\mJd} \,
 \sum_{L_2, \widetilde{m}_s}
  \CG*{J_1}{\mJd}{L_2}{\mJd-\widetilde{m}_s}{S=1}{\widetilde{m}_s} \\
 \null \times  \sum_{L_d = 0, 2}
  \CG{L_d}{\mJd - \widetilde{m}_s}{S=1}{\widetilde{m}_s}{J = 1}{\mJd}
 \int \! \dd k_2 \, k_2^2 \, \widetilde{U}(\pp, k_2, L_1, L_2, J_1, S=1,T_1) \\
 \null \times \int \! \dcostheta \,
 P_{L_2}^{\mJd-\widetilde{m}_s}(\cos\theta) \,
 P_{L_d}^{\mJd-\widetilde{m}_s}\!\big(\cos\thetacprime(k_2, \theta)\big)
 \psi_{L_d}^\lambda\!\left(\sqrt{{k_2}^2 - k_2 q \cos\theta + q^2/4}\right) \,,
\end{multline}
\begin{multline}
 B_1 \equiv \la \phi | \,\widetilde{U}\, J_0 \,\widetilde{U}^\dag \, |
 \psi_i^\lambda \ra = \frac{4}{\pi} \, \sqrt{\frac{2}{\pi}} \,
 \sum_{T_1 = 0,1} \big(G_E ^p + (-1)^{T_1} \, G_E^n\big)
 \sum_{L_1 = 0}^{L_{\rm max}} \big(1 + (-1)^{T_1} (-1)^{L_1}\big) \,
  Y_{L_1 , \mJd - \msf} (\thetacm, \phicm) \\
 \null \times \sum_{J_1 = |L_1 - 1|}^{L+1}
  \CG{L_1}{\mJd-\msf}{S=1}{\msf}{J_1}{\mJd} \,
 \sum_{L_2, \widetilde{m}_s}
  \CG*{J_1}{\mJd}{L_2}{\mJd-\widetilde{m}_s}{S=1}{\widetilde{m}_s} \\
 \null \times \sum_{L_3 = 0}^{L_{\rm max}}
  \CG{L_3}{\mJd-\widetilde{m}_s}{S=1}{\widetilde{m}_s}{J=1}{\mJd}
 \int \! \dd k_2 \, k_2^2 \, \widetilde{U} (\pp, k_2, L_1, L_2, J_1, S = 1,T_1)
 \sum_{L_d = 0, 2} \int \! \dd k_4 \, k_4^2 \, \psi_{L_d}^\lambda (k_4) \, \\
 \null \times \int \! \dcostheta \, P_{L_2}^{\mJd - \widetilde{m}_s}(\cos\theta)
 \,P_{L_3}^{\mJd - \widetilde{m}_s}\!\big(\cos \thetacprime(k_2, \theta)\big) \,
 \widetilde{U}\!\left(
   k_4, \sqrt{{k_2}^2 - k_2 q \cos\theta + q^2/4}, L_d, L_3, J=1, S=1, T=0
 \right) \,.
\end{multline}
In deriving the equations for $B_1$, $B_2$, and $B_3$ we have made use of
the fact that the matrix elements with $J_0$ are twice the matrix elements
with $J_0^-$, \ie, $\mbraket{\phi}{J_0 \, \widetilde{U}^\dag}{\psi_i^\lambda}
= 2 \,\mbraket{\phi}{J_0^- \, \widetilde{U}^\dag}{\psi_i^\lambda}$, and
similarly for $B_2$ and $B_1$ (\cf~Sec.~\ref{sec:calculating_fl}).

Evaluating Eq.~\eqref{eq:A_split_up} involves calculating the individual terms
$A_1$, $A_2$, $A_3$, and $A_4$.  The expressions for $A_4$ and $A_3$ can be
obtained from expressions for $B_2$ and $B_1$, respectively, by replacing
$\widetilde{U}$ with $\widetilde{U}^\dag$.  The $U$-matrices are real.
Therefore, $\widetilde{U}^\dag$ is obtained from $\widetilde{U}$ by
interchanging momentum and angular momentum indices.  The expressions for $A_2$
and $A_1$ are
\begin{multline}
 A_2 \equiv \la \phi | \widetilde{U}^\dag \, \widetilde{U}\, J_0 |
 \psi_i^\lambda \ra  = \frac{4}{\pi} \, \sqrt{\frac{2}{\pi}} \,
 \sum_{T_1 = 0,1} \big(G_E ^p + (-1)^{T_1} \, G_E^n\big)
 \sum_{L_1 = 0}^{L_{\rm max}} \big(1 + (-1)^{T_1} (-1)^{L_1}\big) \,
  Y_{L_1 , \mJd - \msf} (\thetacm, \phicm) \\
 \null \times \sum_{J_1 = |L_1 - 1|}^{L+1}
  \CG{L_1}{\mJd-\msf}{S=1}{\msf}{J_1}{\mJd} \,
 \sum_{L_3, \widetilde{m}_s}
  \CG*{J_1}{\mJd}{L_3}{\mJd-\widetilde{m}_s}{S=1}{\widetilde{m}_s} \\
 \null \times \sum_{L_2 = 0}^{L_{\rm max}} \int \! \dd k_2 \, k_2^2 \,
  \widetilde{U} (k_2, \pp, L_2, L_1, J_1, S=1, T_1)
 \sum_{L_d = 0, 2}
  \CG{L_d}{\mJd-\widetilde{m}_s}{S=1}{\widetilde{m}_s}{J=1}{\mJd} \\
 \null \times \int \! \dd k_3 \, k_3^2 \,
  \widetilde{U} (k_2, k_3, L_2, L_3, J_1, S = 1, T_1)
 \int \! \dcostheta \,P_{L_3}^{\mJd - \widetilde{m}_s}(\cos\theta) \\
 \null \times P_{L_d}^{\mJd - \widetilde{m}_s}\!\big(
  \cos\thetacprime(k_3, \theta)\big) \,
 \psi_{L_d}^\lambda\!\left(\sqrt{{k_3}^2 - k_3 q \cos\theta + q^2/4} \right)
\end{multline}
and
\begin{multline}
 A_1 \equiv \la \phi | \widetilde{U}^\dag \, \widetilde{U}\, J_0 |
 \psi_i^\lambda \ra  = \frac{8}{\pi^2} \,\sqrt{\frac{2}{\pi}} \,
 \sum_{T_1 = 0,1} \big(G_E ^p + (-1)^{T_1} \,G_E^n\big)
 \sum_{L_1 = 0}^{L_{\rm max}} \big(1 + (-1)^{T_1} (-1)^{L_1}\big) \,
  Y_{L_1 , \mJd - \msf} (\thetacm, \phicm) \\
 \null \times \sum_{J_1 = |L_1 - 1|}^{L+1}
  \CG{L_1}{\mJd - \msf}{S=1}{\msf}{J_1}{\mJd} \,
 \sum_{L_3, \widetilde{m}_s}
  \CG*{J_1}{\mJd}{L_3}{\mJd-\widetilde{m}_s}{S=1}{\widetilde{m}_s} \\
 \null \times \sum_{L_4 = 0}^{L_{\rm max}}
  \CG{L_4}{\mJd - \widetilde{m}_s}{S=1}{\widetilde{m}_s}{J=1}{\mJd}
 \sum_{L_2 = 0}^{L_{\rm max}} \int \! \dd k_2 \, k_2^2 \,
  \widetilde{U} (k_2, \pp, L_2, L_1, J_1, S=1, T_1) \\
 \null \times\int \! \dd k_3 \, k_3^2 \,
  \widetilde{U} (k_2, k_3, L_2, L_3, J_1, S=1, T_1)
 \sum_{L_d = 0, 2} \int \! \dd k_5 \, k_5^2 \, \psi_{L_d}^\lambda (k_5)
 \int \! \dcostheta \, P_{L_3}^{\mJd - \widetilde{m}_s}(\cos \theta) \\
 \null \times
 P_{L_4}^{\mJd - \widetilde{m}_s}\!\big(\cos\thetacprime(k_3, \theta)\big)
 \,\widetilde{U}\!\left(
  k_5, \sqrt{{k_3}^2 - k_3 q \cos\theta  + q^2/4}, L_d, L_4, J=1, S=1, T=0
 \right) \,.
\end{multline}

Evaluating the evolved current while including the final-state interactions
involves computing the terms $F_1$, $F_2$, $F_3$, and $F_4$, as indicated in
Eq.~\eqref{eq:F_split_up}.  $F_4$ is obtained from
Eqs.~\eqref{eq:phi_t_g0_J0_minus} and~\eqref{eq:J0_minus_twice_relation} by
replacing the deuteron wave function and the $t$-matrix by their evolved
counterparts.  The expressions for the terms $F_3$, $F_2$, and $F_1$ are then
as follows:
\begin{multline}
 F_3 \equiv \mbraket{\phi}{t_\lambda ^\dag \, G_0^\dag \,
 J_0 \, \widetilde{U}^\dag}{\psi_i^\lambda}
 = \frac{4}{\pi} \, \sqrt{\frac{2}{\pi}} \, \frac{M}{\hbar c}
 \int \! \frac{\dd k_2 \, k_2^2}{(\pp + k_2)(\pp - k_2 - i \epsilon)}
 \sum_{T_1 = 0,1} \big(G_E ^p + (-1)^{T_1} \,G_E^n\big)
 \sum_{L_1 = 0}^{L_{\rm max}} \big(1 + (-1)^{T_1} (-1)^{L_1}\big) \\
 \null \times Y_{L_1 , \mJd - \msf}(\thetacm, \phicm)
 \sum_{J_1 = |L_1 - 1|}^{L+1}
  \CG{L_1}{\mJd - \msf}{S=1}{\msf}{J_1}{\mJd}
 \sum_{L_2 = 0}^{L_{\rm max}}
  t^\ast_\lambda(k_2, \pp, L_2, L_1, J_1, S=1, T_1) \\
 \null \times \sum_{\widetilde{m}_s = -1}^1
  \CG*{J_1}{\mJd}{L_2}{\mJd-\widetilde{m}_s}{S=1}{\widetilde{m}_s}
 \sum_{L_3 = 0}^{L_{\rm max}}
  \CG{L_3}{\mJd - \widetilde{m}_s}{S = 1}{\widetilde{m}_s}{J=1}{\mJd}
 \int \! \dcostheta \, P_{L_2}^{\mJd - \widetilde{m}_s}(\cos \theta) \\
 \null \times
 P_{L_3}^{\mJd - \widetilde{m}_s}\!\big(\cos \thetacprime(k_2, \theta)\big)
 \int \! \dd k_5 \, k_5^2 \sum_{L_d = 0, 2}
 \widetilde{U}\!\left(
  k_5, \sqrt{{k_2}^2 - k_2 q \cos\theta + q^2/4}, L_d, L_3, J=1, S=1, T=0
 \right) \psi_{L_d}^\lambda (k_5) \,,
\label{eq:F3}
\end{multline}
\begin{multline}
 F_2 \equiv \mbraket{\phi}{t_\lambda ^\dag \, G_0^\dag \, \widetilde{U} \,
 J_0}{\psi_i^\lambda}
 = \frac{4}{\pi} \, \sqrt{\frac{2}{\pi}} \,\frac{M}{\hbar c}
 \int \! \frac{\dd k_2 \, k_2^2}{(\pp + k_2)(\pp - k_2 - i \epsilon)}
 \sum_{T_1 = 0,1} \big(G_E ^p + (-1)^{T_1} \,G_E^n\big)
 \sum_{L_1 = 0}^{L_{\rm max}} \big(1 + (-1)^{T_1} (-1)^{L_1}\big) \\
 \null \times Y_{L_1 , \mJd - \msf} (\thetacm, \phicm)
 \sum_{J_1 = |L_1 - 1|}^{L+1}
  \CG{L_1}{\mJd - \msf}{S=1}{\msf}{J_1}{\mJd}
 \sum_{L_2 = 0}^{L_{\rm max}}
  t^\ast_\lambda(k_2, \pp, L_2, L_1, J_1, S = 1,T_1) \\
 \null \times \sum_{L_3 = 0}^{L_{\rm max}} \int \! \dd k_4 \, k_4^2 \,
 \widetilde{U}(k_2, k_4, L_2, L_3, J_1, S=1, T_1)
 \sum_{\widetilde{m}_s = -1}^1
  \CG*{J_1}{\mJd}{L_3}{\mJd - \widetilde{m}_s}{S = 1}{\widetilde{m}_s} \\
 \null \times \sum_{L_d = 0, 2}
  \CG{L_d}{\mJd - \widetilde{m}_s}{S=1}{\widetilde{m}_s}{J = 1}{\mJd}
 \int \! \dcostheta \, P_{L_3}^{\mJd - \widetilde{m}_s}(\cos \theta) \\
 \null \times
 P_{L_d}^{\mJd - \widetilde{m}_s} \big(\cos \thetacprime(k_4, \theta)\big) \,
 \psi_{L_d}^\lambda\!\left(\sqrt{{k_4}^2 - k_4 q \cos\theta + q^2/4}\right) \,,
\label{eq:F2}
\end{multline}
\begin{multline}
 F_1 \equiv \mbraket{\phi}{t_\lambda ^\dag \, G_0^\dag \, \widetilde{U} \,
 J_0 \, \widetilde{U}^\dag}{\psi_i^\lambda}
 = \frac{8}{\pi^2} \, \sqrt{\frac{2}{\pi}} \, \frac{M}{\hbar c}
 \int \! \frac{\dd k_2 \, k_2^2}{(\pp + k_2)(\pp - k_2 - i \epsilon)}
 \sum_{T_1 = 0,1} \big(G_E ^p + (-1)^{T_1} \,G_E^n\big)
 \sum_{L_1 = 0}^{L_{\rm max}} \big(1 + (-1)^{T_1} (-1)^{L_1}\big) \\
 \null \times Y_{L_1 , \mJd - \msf} (\thetacm, \phicm)
 \sum_{J_1 = |L_1 - 1|}^{L+1}
  \CG{L_1}{\mJd - \msf}{S=1}{\msf}{J_1}{\mJd}
 \sum_{L_2 = 0}^{L_{\rm max}}
  t^\ast_\lambda(k_2, \pp, L_2, L_1, J_1, S=1, T_1) \\
 \null\times \sum_{L_3 = 0}^{L_{\rm max}} \, \sum_{\widetilde{m}_s = -1}^1
  \CG*{J_1}{\mJd}{L_3}{\mJd - \widetilde{m}_s}{S = 1}{\widetilde{m}_s}
 \sum_{L_4 = 0}^{L_{\rm max}}
  \CG{L_4}{\mJd - \widetilde{m}_s}{S = 1}{\widetilde{m}_s}{J=1}{\mJd} \\
 \null \times \int \! \dd k_4 \, k_4^2 \,
  \widetilde{U}(k_2, k_4, L_2, L_3, J_1, S=1, T_1)
 \int \! \dcostheta \, P_{L_3}^{\mJd - \widetilde{m}_s}(\cos \theta) \,
 P_{L_4}^{\mJd - \widetilde{m}_s}\!\big(\cos \thetacprime(k_4, \theta)\big) \\
 \null \times \int \! \dd k_6 \, k_6^2 \sum_{L_d = 0, 2}
 \widetilde{U}\!\left(
  k_6, \sqrt{{k_4}^2 - k_4 q \cos\theta + q^2/4}, L_d, L_4, J=1, S=1, T=0
 \right) \psi_{L_d}^\lambda (k_6) \,.
\label{eq:F1}
\end{multline}

\end{widetext}

\section{Evolved final state}
\label{sec:finalstate}

The interacting final neutron-proton state $\ket{\psi_f}$ as defined in
Eq.~\eqref{eq:psi_f_def} is the formal solution of the Lippmann--Schwinger (LS)
equation for the scattering wave function,
\begin{spliteq}
 \ket{\psi_f}
 &= \ket{\phi} + G_0 (E^\prime) \, V \ket{\psi_f} \\
 &= \ket{\phi} + G_0 (E^\prime) \, t(E^\prime) \ket{\phi} \,.
 \label{eq:LS-psi}
\end{spliteq}
The $t$-matrix, in turn, is defined by the LS equation
\begin{equation}
 t(E^\prime) = V + V \, G_0(E') \, t(E') \,.
\label{eq:LS-t}
\end{equation}
The subsitution $E' \to E' + \ii\epsilon$ and the limit $\epsilon\to0$
are implied to select outgoing boundary conditions.  We want to show now
that the SRG-evolved final state can be obtained directly by using the solution
$t^\lambda$ of Eq.~\eqref{eq:LS-t} with $V \to V_\lambda$ in the second line of
Eq.~\eqref{eq:LS-psi}, which is the same as Eq.~\eqref{eq:psi_f_def} in
Sec.~\ref{sec:formalism}, \ie,
\begin{equation}
 U_\lambda \ket{\psi_f} = \ket{\psi_f^\lambda} \,,
\label{eq:U-psi-final}
\end{equation}
where
\begin{equation}
 \ket{\psi_f^\lambda} = \ket{\phi} +
 G_0 (E^\prime) \, t_\lambda(E^\prime) \ket{\phi} \,.
\end{equation}
In this section, we suppress all spin and isospin degrees of freedom, and only
denote the (arbitrary) energy parameter as $E^\prime$ for consistency with
Sec.~\ref{sec:formalism}.

First, it is important to recall that by definition the free Hamiltonian
$H_0$ does not evolve, so that for $H = H_0 + V$ we have
\begin{equation}
 H_\lambda = U_\lambda \;\! H \, U_\lambda^\dagger
 \equiv H_0 + V_\lambda \,.
\label{eq:H-lambda}
\end{equation}
In other words, the evolved potential $V_\lambda$ is defined such that it
absorbs the evolution of the initial free Hamiltonian (kinetic energy) as well.

In order to prove Eq.~\eqref{eq:U-psi-final}, it is convenient to consider the
evolved and unevolved full Green's functions $G_\lambda(E^\prime)$ and
$G(E^\prime)$, defined via
\begin{subalign}[eq:G-inv]
 G_\lambda(z)^{-1} &= z - H_\lambda = G_0^{-1}(z)^{-1} - V_\lambda \,,
 \label{eq:G-lambda-inv} \\
 G(z)^{-1} &= z - H_{\phantom{\lambda}} = G_0^{-1}(z)^{-1} - V \,.
 \label{eq:G-bare-inv}
\end{subalign}
Here, $G_0^{-1}(z)^{-1} = z - H_0$ is the free Green's function (which does not
change under the SRG evolution because $H_0$ does not), and $z$ is an arbitrary
complex energy parameter that is set to $E^\prime + \ii\epsilon$ to recover the
physically relevant case.  The Green's functions can be expressed in terms of
the $t$-matrix as
\begin{equation}
 G(z) = G_0(z) + G_0(z) \, t(z) \, G_0(z) \,,
\end{equation}
and analogously for the evolved version.  Furthermore, the Green's functions
can be written in their spectral representations
\begin{subalign}[eq:G-spectral]
 G_\lambda(z)^{-1} &\simeq \int\!\dd^3 k
 \, \frac{\ket{\psi_f^\lambda(k)}\bra{\psi_f^\lambda(k)}}{z - k^2/M}
 + \text{bound states} \,,
 \label{eq:G-lambda-spectral} \\
 G(z)^{-1} &\simeq \int\!\dd^3 k
 \, \frac{\ket{\psi_f(k)}\bra{\psi_f(k)}}{z - k^2/M}
 + \text{bound states} \,.
 \label{eq:G-bare-spectral}
\end{subalign}
Here, $\ket{\psi_f^{(\lambda)}(k)}$ denotes the (evolved) continuum states with
momentum $k$, and we have $\ket{\psi_f^{(\lambda)}}
= \ket{\psi_f^{(\lambda)}(\sqrt{M E^\prime})}$

From Eqs.~\eqref{eq:H-lambda} and~\eqref{eq:G-inv} it now follows that
\begin{multline}
 G_\lambda(z)^{-1} = z - H_\lambda
 = z - U_\lambda \;\! H \, U_\lambda^\dagger \\
 = U_\lambda (z - H) \:\! U_\lambda^\dagger
 = U_\lambda \, G(z)^{-1} \, U_\lambda^\dagger \,.
\end{multline}
Combining this with Eqs.~\eqref{eq:G-spectral} and matching residues at $z =
E^\prime + \ii\epsilon$, we find that indeed $\ket{\psi_f^{\lambda}} =
U_\lambda \ket{\psi_f}$, as stated in Eq.~\eqref{eq:U-psi-final}.

\end{document}